\documentclass[aps,prx,showpacs,floatfix,twocolumn,superscriptaddress,longbibliography]{revtex4-2}

\usepackage{epsf}
\usepackage{epsfig}
\usepackage{amsmath}
\usepackage{amsfonts}
\usepackage{amssymb}
\usepackage{graphicx}
\usepackage{multirow}
\usepackage{graphicx}
\usepackage[table,xcdraw,svgnames]{xcolor}
\usepackage{diagbox}

\usepackage{epsfig}
\usepackage{color}
\usepackage{bm}
\usepackage[yyyymmdd]{datetime}
\usepackage{dashrule}
\usepackage{ulem}
\usepackage{lineno}
\usepackage{pifont}
\usepackage{array}
\usepackage[margin=0.7in]{geometry}

\newcommand{\jl}[1]{\textcolor{black}{#1}}

\newcommand{\cmark}{\textcolor{black}{\ding{51}}}

\newcommand{\xmark}{\textcolor{black}{\ding{55}}}
\newcommand{\xrmark}{\textcolor{red}{\ding{55}}}
\newcommand{\xbmark}{\textcolor{blue}{\ding{55}}}

\begin{document}
    \title{Single- and Multimagnon Dynamics in Antiferromagnetic $\alpha$-Fe$_2$O$_3$ Thin Films}
    \author{Jiemin Li}
    \email{jli1@bnl.gov}
    \affiliation{National Synchrotron Light Source II, Brookhaven National Laboratory, Upton, NY 11973, USA.}
    \author{Yanhong Gu}
    \affiliation{National Synchrotron Light Source II, Brookhaven National Laboratory, Upton, NY 11973, USA.}
    \author{Yoshihiro Takahashi}
    \affiliation{Department of Physics and Electronics, Osaka Metropolitan University, 1-1 Gakuen-cho, Nakaku, Sakai, Osaka 599-8531, Japan.}
    \author{Keisuke Higashi}
    \affiliation{Department of Physics and Electronics, Osaka Metropolitan University, 1-1 Gakuen-cho, Nakaku, Sakai, Osaka 599-8531, Japan.}
    \author{Taehun Kim}
    \affiliation{National Synchrotron Light Source II, Brookhaven National Laboratory, Upton, NY 11973, USA.}
    \author{Yang Cheng}
    \affiliation{Department of Physics, The Ohio State University, Columbus, OH 43210, USA.}
    \author{Fengyuan Yang}
    \affiliation{Department of Physics, The Ohio State University, Columbus, OH 43210, USA.}
    \author{Jan Kune\v{s}}
    \affiliation{Institute of Solid State Physics, TU Wien, 1040 Vienna, Austria.}
    \author{Jonathan Pelliciari}
    \affiliation{National Synchrotron Light Source II, Brookhaven National Laboratory, Upton, NY 11973, USA.}
    \author{Atsushi Hariki}
    \affiliation{Department of Physics and Electronics, Osaka Metropolitan University, 1-1 Gakuen-cho, Nakaku, Sakai, Osaka 599-8531, Japan.}
    \author{Valentina Bisogni}
    \email{bisogni@bnl.gov}
    \affiliation{National Synchrotron Light Source II, Brookhaven National Laboratory, Upton, NY 11973, USA.}


    \begin{abstract}
        Understanding the spin dynamics in antiferromagnetic (AFM) thin films is fundamental for designing novel devices based on AFM magnon transport. Here, we study the magnon dynamics in thin films of AFM $S=\frac{5}{2}$ $\alpha$-Fe$_2$O$_3$ by combining resonant inelastic x-ray scattering, Anderson impurity model plus dynamical mean-field theory, and Heisenberg spin model. Below 100~meV, we observe the thickness-independent (down to 15~nm) acoustic single-magnon mode. At higher energies ($100-500$~meV), an unexpected sequence of equally spaced, optical modes is resolved and ascribed to $\Delta S_z=1, 2, 3, 4$, and 5 magnetic excitations corresponding to multiple, noninteracting magnons. Our study unveils the energy, character, and momentum dependence of single and multimagnons in $\alpha$-Fe$_2$O$_3$ thin films, with impact on AFM magnon transport and its related phenomena. 
        From a broader perspective, we generalize the use of $L$-edge resonant inelastic x-ray scattering as a multispin-excitation probe up to $\Delta S_z= 2S$. Our analysis identifies the spin-orbital mixing in the valence shell as the key element for accessing excitations beyond $\Delta S_z= 1$, and up to, e.g., $\Delta S_z=5$. \jl{At the same time, we elucidate the novel origin of the spin excitations beyond the $\Delta S_z= 2$, emphasizing the key role played by the crystal lattice as a reservoir of angular momentum that complements the quanta carried by the absorbed and emitted photons}.
    \end{abstract}

    \maketitle

    \section{Introduction}
    The never-ending demand for faster and low-power devices is stimulating the development of novel electronics such as 
    antiferromagnetic spintronics \cite{Jungwirth16,Baltz18}. In this context, insulating antiferromagnets
    are highly attractive as they support novel spin-transport phenomena conveyed by AFM magnons \cite{Wang2014,Wu2016}, without involving moving charges. As a result, renewed interest is now focused on well-known AFM materials, such as $\alpha$-Fe$_2$O$_3$ (hematite) which offers functional opportunities including long-distance spin transport \cite{Lebrun18,Lebrun20,Han20} and electrical switching of the N\'{e}el order \cite{Cheng19,Zhang19,Cheng20,Cogulu21}.
    Hence, to progress in 
    AFM spintronics, magnon transport, and its underlying microscopic phenomena, it is crucial to understand the whole magnon spectrum, i.e., single and multimagnons, in thin-film form as used for devices. Indeed, thin films often present deviations in the spin \textbf{\textit{q}}-dynamics with respect to the bulk, owing to strain, confinement, electronic band reconstruction, etc. \cite{Meyers19, Ivashko19,Pelliciari21_2,Pelliciari21}. 
    
    Only limited information is, however, available on magnon 
    dynamics in thin films, mostly due to lack of suitable experimental probes.
    Thanks to dramatic improvements in the energy resolution, resonant inelastic x-ray scattering (RIXS) is emerging as a powerful tool for the study of spin dynamics in magnetic materials \cite{Braicovich10, Ament11, Fabbris17, Betto17, Elnaggar19, Li21, Nag20, Betto21, Martinelli22, Gu22}, and in thin films \cite{Dean12, Brookes20, Pelliciari21_2}. 
    Concomitantly, relevant progress in the theoretical description of the RIXS process contributed to elucidate the complex magnetic interactions behind the RIXS cross section~\cite{Ament11, Haverkort10, Zimmermann18, Wang19, Hariki20, Gilmore21}. As an example, a unique sensitivity to multispin excitations up to $\Delta S_z$=2 was revealed \cite{Ghiringhelli09, Bisogni14, Nag20, Schlappa18, Nag22}, contrary  to conventional magnon probes. 
    Therefore, this novel approach promises to bridge the present gap in the study of single and multimagnons, in thin film and bulk materials.
    
    Here, we focus on $S=\frac{5}{2}$ $\alpha$-Fe$_2$O$_3$ aiming to investigate the magnon modes and their {\textit{\textbf{q}}}-dynamics in 30- and 15-nm thin films by using high-resolution Fe $L_3$-edge RIXS. Below 100~meV, the dispersing single-magnon mode revealed in the two thin films perfectly match each other as well as the same mode in the bulk \cite{Samuelsen69,Samuelsen70, Shull51}. At higher energies, our data instead 
    displays unexpected, equally spaced peaks in the approximately ($100-500$)-meV range with negligible {\textit{\textbf{q}}}-dependence. Combining
    Anderson impurity model (AIM) built on local density approximation (LDA) plus dynamical mean-field theory (DMFT) and Heisenberg model analysis, we demonstrate these modes to be magnetic and noninteracting, corresponding to $\Delta S_z=1, 2, 3, 4$, and 5 multimagnon excitations.  Overall, our study provides information on the fundamental magnetic modes in hematite thin films, which are of relevance, e.g., for magnon transport and pumping phenomena. At the same time, we highlight $L$-edge RIXS as an effective method for accessing multimagnon excitations. Thanks to an analysis of the spin character during the intermediate state, we identify the spin-orbital mixing in the valence shell as the key element for accessing magnons beyond $\Delta S_z= 1$ and up to $\Delta S_z=2S$. We also demonstrate that the lattice-breaking spherical symmetry plays a crucial role in accessing $\Delta S_z>2$, effectively acting as a reservoir for angular momentum.

        \begin{figure*}
            \centering
            {\includegraphics{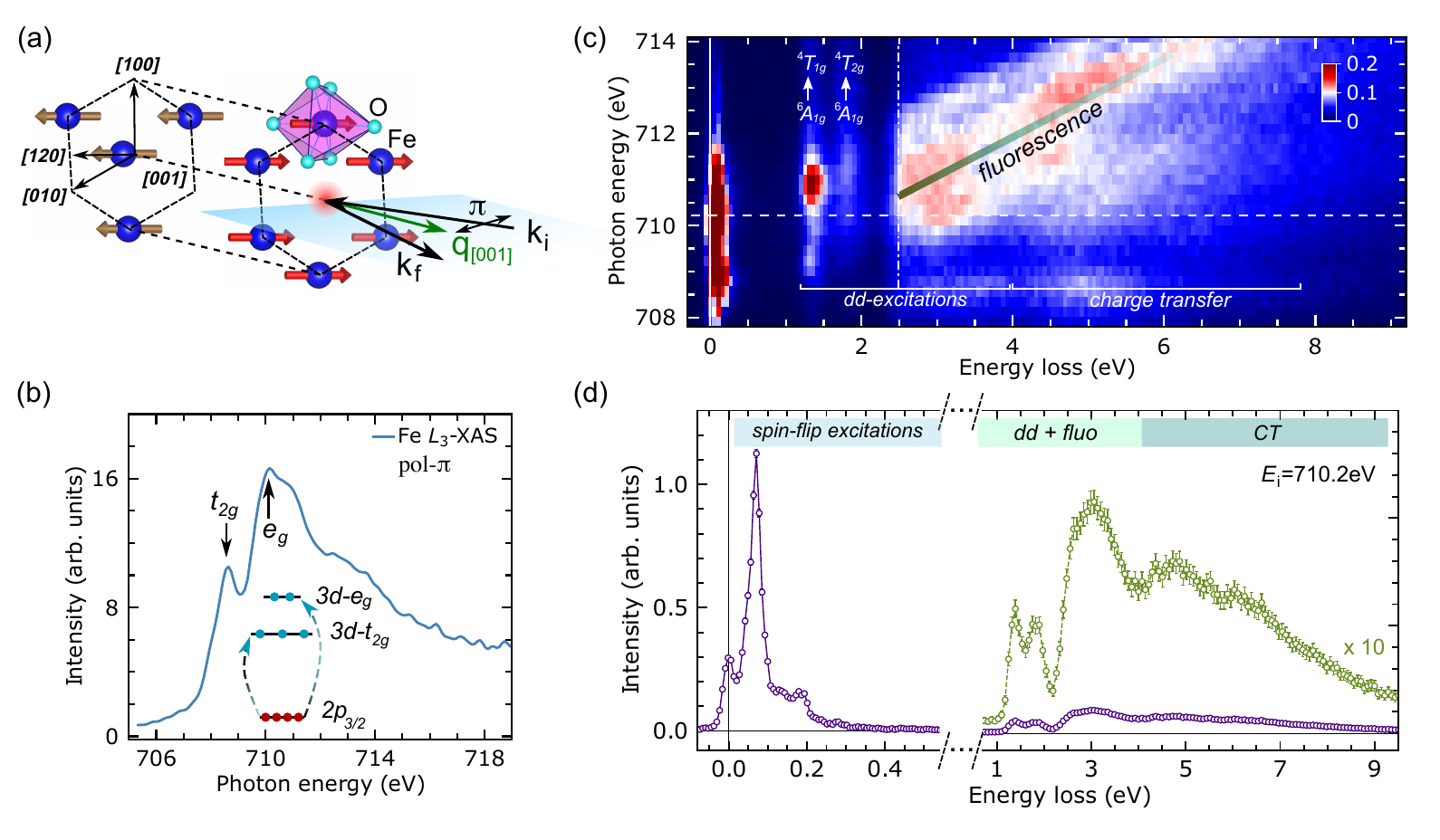}}
            \caption{Scattering geometry and experimental data for 30-nm $\alpha$-Fe$_2$O$_3$ film. (a) Crystal and magnetic structure of $\alpha$-Fe$_2$O$_3$. The blue shaded area indicates the scattering plane defined by the [001] and [120] axes of the hexagonal unit cell. The black arrows labeled $k_i$ ($k_f$) represent the incident (scattered) x rays, while the double arrow refers to the $\pi$ polarization. The momentum transfer $\textit{\textbf{q}}$ (green arrow) is along the [001] direction. (b) Fe $L_3$-edge XAS measured in the PFY mode using the RIXS spectrometer at Fe $L_3$ edge, $\theta=33.5^{\circ}$ and $2\theta=67^{\circ}$. The inset depicts the related Fe $L_3$ XAS process. \textcolor{black}{XAS in total fluorescence or electron yield mode cannot be successfully measured in this sample due to technical difficulties.} (c) RIXS energy dependence across the Fe $L_3$ edge at $\theta=20^{\circ}$ and 2$\theta=150^{\circ}$, and $\pi$-polarized x rays. The horizontal dashed line indicates the incident photon energy 710.2~eV used for RIXS measurements, while the vertical dot-dashed line specifies the onset of fluorescence. (d) Overview of a typical high-resolution RIXS spectrum in purple color ($\theta=45^{\circ}$, $2\theta=90^{\circ}$ and $\textit{\textbf{q}}$=0.51~\AA$^{-1}$). The high-energy-loss excitations are multiplied by a factor of 10 in the green colored spectrum to ease the visualization of the $dd$, charge transfer, and fluorescence fine structure. The error bars are defined assuming a Poisson distribution of the centroided, single-photon events. 
            }
            \label{Fig1}
        \end{figure*}

        \section{Experimental details}
        High-quality crystalline thin films of $\alpha$-Fe$_2$O$_3$ with 30 and 15~nm thickness were grown fully relaxed on Al$_2$O$_3$~\cite{Cheng20}, realizing the typical hematite corundum structure (see Sec. I of the Supplemental Material~\cite{sm} for further details). The spin momentum $S=\frac{5}{2}$, deriving from the 3$d^5$ high-spin electronic configuration of the Fe$^{3+}$ ions [see inset of Fig.~\ref{Fig1}(b)], lies in the (001) plane and stacks antiferromagnetically along the $c$ axis of the hexagonal cell; see Fig.~\ref{Fig1}(a).
        
        Fe $L_3$-edge x-ray absorption spectroscopy (XAS) and RIXS measurements are performed at the SIX 2-ID beamline of the National Synchrotron Light Source II~\cite{Dvorak16}. The RIXS experiment is conducted with an energy resolution of $\Delta E\sim$ 23~meV (full width at half maximum) at the Fe $L_3$ edge. Linear-horizontal ($\pi$) polarization of the incident light is used to minimize the elastic scattering in the RIXS spectra. The films are oriented with the [120] and [001] axes lying in the scattering plane; see Fig.~\ref{Fig1}(a). The temperature is kept at $T$=100~K. The momentum $\textit{\textbf{q}}$ transferred during the RIXS measurements is parallel to the [001] axis  and defined in absolute unit \AA$^{-1}$.

        \section{Magnetic excitations of $\alpha$-F\lowercase{e}$_2$O$_3$ films}
        Figure~\ref{Fig1}(b) shows the Fe $L_3$-edge XAS of 30-nm $\alpha$-Fe$_2$O$_3$ film measured in partial fluorescence yield (PFY) mode. As the octahedral crystal field splits the Fe 3$d$ orbitals into $t_{2g}$ and $e_{g}$ levels, the XAS peak at 708.8~eV (710.2~eV) corresponds to the transition \textcolor{black}{from the 
        2$p_{3/2}$ orbitals to the 
        $t_{2g}$ ($e_{g}$)} ones~\cite{Kuiper93}, as illustrated in the inset of Fig.~\ref{Fig1}(b).
        
        Figure~\ref{Fig1}(c) presents the energy dependence of the RIXS spectra across the Fe $L_3$ edge, in the same sample. The color map reveals pronounced excitations in two well-separated energy-loss windows, respectively, below 500~meV and above 1~eV. In the latter case, we observe interorbital, Raman-like, $dd$ excitations in the energy range of 1-4~eV, with two main branches emerging at approximately 1.4 eV and 1.9 eV. These $dd$ excitations correspond to $^6A_{1g}\rightarrow^4T_{1g}$ and $^6A_{1g}\rightarrow^4T_{2g}$ transitions, flipping both spin ($S=\frac{5}{2} \rightarrow \frac{3}{2}$) and orbital ($e_g \rightarrow t_{2g}$) respectively. At higher energies, weak charge-transfer excitations involving the electron hopping between Fe 3$d$ and O 2$p$ orbitals start from approximately 4-eV energy loss and extend up to 10-eV. Furthermore, when the photon energy approaches the Fe $e_{g}$ resonance [dashed line in Fig.~\ref{Fig1}(c)],  a strong fluorescence emerges around 2.5~eV, shifting linearly in energy loss as the incident photon energy increases. Overall, the broad high-energy excitations discussed above are consistent with previous low-resolution RIXS studies of $\alpha$-Fe$_2$O$_3$~\cite{Miyawaki17, Hariki20, Ellis22}. 
        
        To introduce the low-energy excitations (below 500~meV), we display in Fig.~\ref{Fig1}(d) a representative high-resolution RIXS spectrum recorded at the Fe $e_g$ resonance as it maximizes the intensity of the excitations below 100~meV (see Fig.~S1 and Sec.~II in the Supplementary Material~\cite{sm}). The so far unresolved low-energy range of the hematite RIXS spectrum reveals multiple sharp excitations with the mode at approximately 100~meV appearing as resolution limited and dominating the overall spectral weight. Similar low-energy features with identical energies and intensities are observed in the 15-nm $\alpha$-Fe$_2$O$_3$ film in the same scattering conditions. The corresponding results are presented in Fig.~S2 and Sec.~III of the Supplementary Material \cite{sm}. 

        \begin{figure}[t]
            \includegraphics[width=1.00\columnwidth]{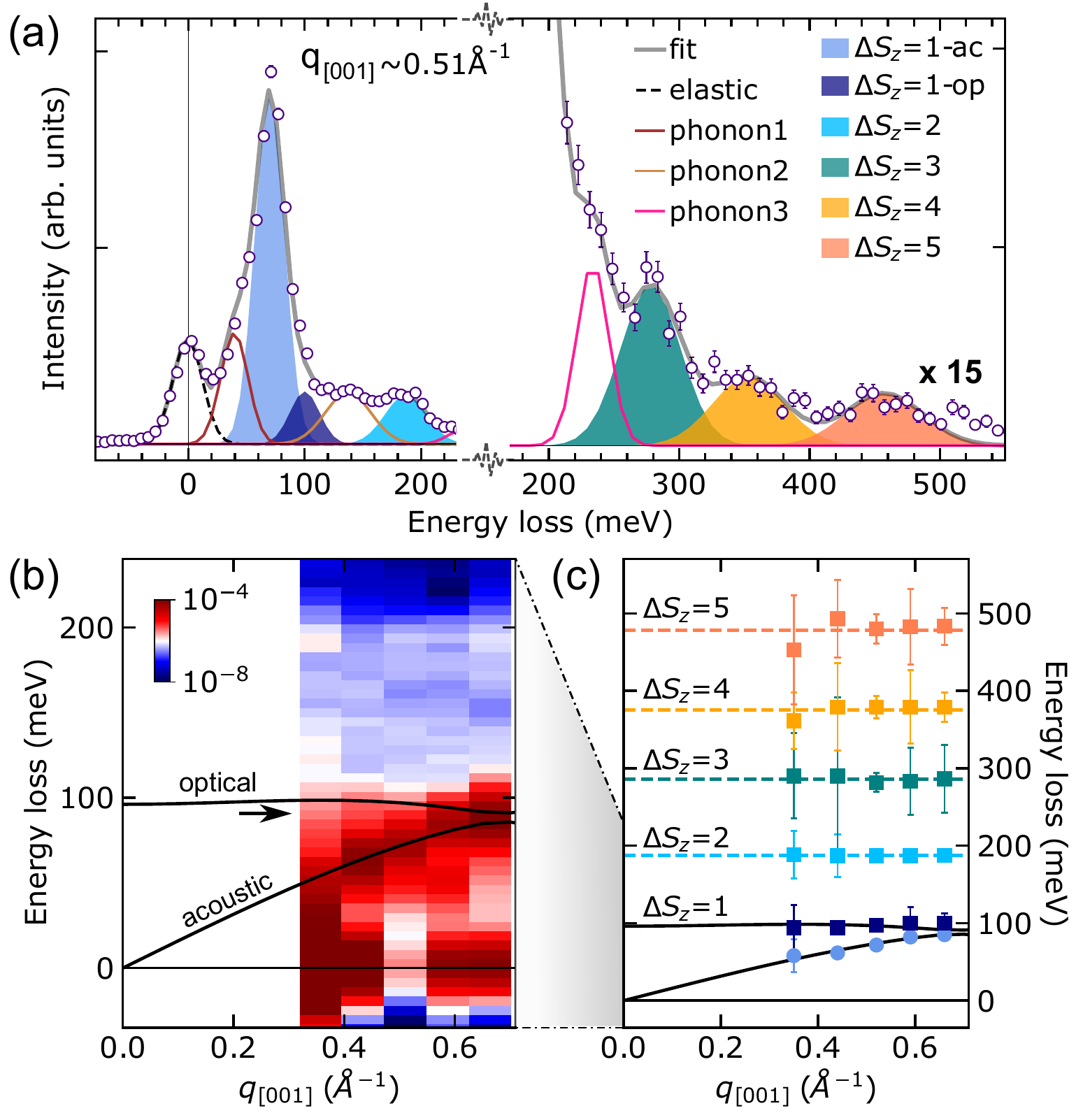}
            \caption{ Momentum dependence of low-energy excitations in 30-nm $\alpha$-Fe$_2$O$_3$ film. (a) Englarged view of the selected RIXS spectrum in Fig.~\ref{Fig1}(c) at  \textbf{\textit{q}}=0.51~\AA$^{-1}$ (open dots). The low-energy excitations up to 500~meV are fitted with ten Gaussians, as explained in the main text. The gray solid line is the sum of the fitted components. (b) RIXS color map (in logarithmic scale) as a function of \textbf{\textit{q}}. The spectra are normalized to the integrated spectral weight in the region 0.6~eV-8.0~eV. The black lines display the calculated single-magnon acoustic and optical branches for $\alpha$-Fe$_2$O$_3$ single crystal~\cite{Samuelsen70}. (c) Fitting results in  terms of peak position versus \textbf{\textit{q}} for the six unconstrained solid-filled Gaussians. The black lines are the same as in (b) and the color code is the same as in (a). The fit error bar is defined in Sec. V of the Supplementary Material~\cite{sm}.
            }
            \label{Fig2}
        \end{figure}

        As the main scope of this work is about the nature of these low-energy excitations, we present in Fig.~\ref{Fig2}(a) a closer view of the RIXS spectrum introduced in Fig.~\ref{Fig1}(d). A long sequence of excitations extending up to 500~meV is observed. Note that the intensity of the region between 200 and 500 meV is expanded by a factor of 15 for clarity.
        While it is known that magnon modes below 100~meV \cite{Samuelsen70} and (multi) phonon modes up to 200 meV \cite{Massey90, Azuma05} exist in hematite, the richness of the excitations observed in the RIXS data up to 500 meV is unprecedented. To scrutinize the origin of these modes, we conduct a momentum dependent study, with $\textit{\textbf{q}}$ along the out-of-plane [001] direction. The result is displayed in Fig.~\ref{Fig2}(b), for energy losses up to 220 meV. A peak is observed dispersing upward in energy loss for increasing \textbf{\textit{q}} and reaching approximately 100~meV at the zone boundary. Additional peaks with small or no momentum dependence are visible in the (100-200)~meV region of Fig.~\ref{Fig2}(b). Because of the weaker intensity of the latter excitations with respect to the ones below 100 meV, we fit all contributions in the RIXS spectra 
        \textcolor{black}{within the (-100, 570)-meV range} to extract quantitative information about their \textbf{\textit{q}}-dependence. In Fig.~\ref{Fig2}(a) we introduce our fitting model based on ten Gaussian profiles, and we assign one peak to the elastic line at 0 meV (dashed line), three peaks to a known phonon and multiphonon contributions (solid lines) from Refs. \cite{Chamritski05, Massey90, Azuma05} (see also Sec. IV in the Supplementary Material~\cite{sm}), and additional six peaks (area-filled profiles) \textcolor{black}{to all the other excitations extending up to 500 meV}. For the details of our fitting analysis \textcolor{black}{and the implemented constraints}, refer to Sec.~V of the Supplementary Material~\cite{sm}. In Fig.~\ref{Fig2}(c) we report a summary of the fitted energies for the six area-filled peaks. Except the lowest-energy excitation behaving as an acoustic mode, all the other excitations are optical modes: Interestingly, their \textbf{\textit{q}}-averaged energies appear to follow a harmonic sequence with respect to the first optical mode at $E_0$, i.e., $E_0$ ($\sim$~97$\pm$1 meV),  2$E_0$ ($\sim$~188$\pm$4 meV), 3$E_0$ ($\sim$~286$\pm$15 meV), 4$E_0$ ($\sim$~376$\pm$17 meV) and 5$E_0$ ($\sim$~478$\pm$19 meV). 
        
        From inelastic-neutron-scattering studies \cite{Samuelsen69, Samuelsen70}, the single magnon ($\Delta S_z=1$ excitation) in $\alpha$-Fe$_2$O$_3$ single crystal is known to exhibit both an acoustic and an optical branch. In Figs.~\ref{Fig2}(b) and 2(c), we overlay to the RIXS data the single-magnon dispersion calculated following Ref.~\cite{Samuelsen70}. A remarkable consistency exists between the calculated dispersion of the $\Delta S_z=1$ branches and the excitations probed by RIXS below 100~meV. On one hand, this important observation reveals the magnetic nature of the first two excitations in Fig.~\ref{Fig2}(c). On the other hand, this result shows that the out-of-plane single magnon has the same energy either in the bulk or in the thin films (30 and 15~nm thick). This suggests that the single magnon in hematite is not subject to thickness-related phenomena or to electronic band reconstruction  acting on the confinement direction \cite{Pelliciari21_2, Pelliciari21}. Finally, we propose that the observed harmonic sequence of optical modes at 2$E_0$, 3$E_0$, 4$E_0$, and 5$E_0$ energies may share the same magnetic origin as the fundamental mode $\Delta S_z=1$ at $E_0$ energy.


        \section{Calculations and discussion}
        To assess the nature of the low-energy harmoniclike sequence of excitations,  
        we use the LDA+DMFT AIM, a recently introduced method to simulate the RIXS spectra of correlated transition-metal oxides~\cite{Hariki18,Hariki20,Winder20,Higashi21}. The LDA+DMFT calculations are performed for the valence electrons (Fe 3$d$ and O 2$p$ orbitals are included) in the experimental crystal structure~\cite{Georges96,Kotliar06,Kunes09,Kunes09r}.
        In addition to the {\it ab initio} band structure, three parameters are required
        :~Hubbard $U$, Hund's coupling $J$, and charge-transfer energy $\Delta_{\rm CT}$~\cite{fn1,Kunes09,Hariki20,Winder20}. These are determined to be $U$=~6.5~eV, $\Delta_{\rm CT}$=~2.4~eV and $J$=~1.0~eV by matching the experimental RIXS high-energy excitations (above 1~eV) with LDA+DMFT AIM calculations, as detailed in the Sec.~VI of the Supplementary Material~\cite{sm}.

        Figure~~\ref{Fig3}(a) shows the computed RIXS 
        intensities evaluated using the Kramers-Heisenberg formula (see Sec.~VI of the Supplementary Material~\cite{sm}), which implements the dipole transitions for the photon absorption and emission.
        The spectrum (black line) exhibits an intense peak at $E_{\rm SF}=66$~meV, followed by four equally spaced peaks at 2$E_{\rm SF}$, 3$E_{\rm SF}$, 4$E_{\rm SF}$, and 5$E_{\rm SF}$ with decreasing intensities, similar to the experimental observation represented by the fitted Gaussian components [color-filled areas in Fig.~\ref{Fig3}(a)]. 
        Analyzing the ground-state and final-state wave functions (see the Table~III in the Supplementary Material~\cite{sm}), we identify the calculated peaks as 
        $\Delta S_z=1,~2,~3,~4,$ and 5
        multispin-flip excitations with increasing energy; see the diagram in Fig.~\ref{Fig3}(b).
        Although there is no analogy of the acoustic magnon branch in AIM, the local spin-flip excitations captured by the Anderson impurity model coincide with the real energy of the flat optical magnon branch in a 3D antiferromagnet.
        
        The LDA+DMFT AIM, however, appears to underestimate the peak positions of  $\Delta S_z=1,~2,~3,~4,$ and 5 excitations by a factor of approximately 1.4 with respect to the experiment; see the different scales used for the bottom (calculation) and top (experiment) axes in Fig.~\ref{Fig3}(a). 
        This result reflects an underestimation of the effective interatomic spin exchange by the present LDA+DMFT model. 
        \jl{A more accurate description of the effective exchange parameters may require 
        inclusion of interaction on the oxygen sites~\cite{Logemann17}, which is beyond the scope of this work.} \jl{Furthermore, we note that the calculation underestimates the spectral weight of $\Delta S_z=3,~4,~5$ excitations by  a factor of $\sim$~4 relative to that of $\Delta S_z=1, 2$ excitations. Possible causes for this are discussed in Sec. VI of~\cite{sm}.}

        \begin{figure*}
            \includegraphics{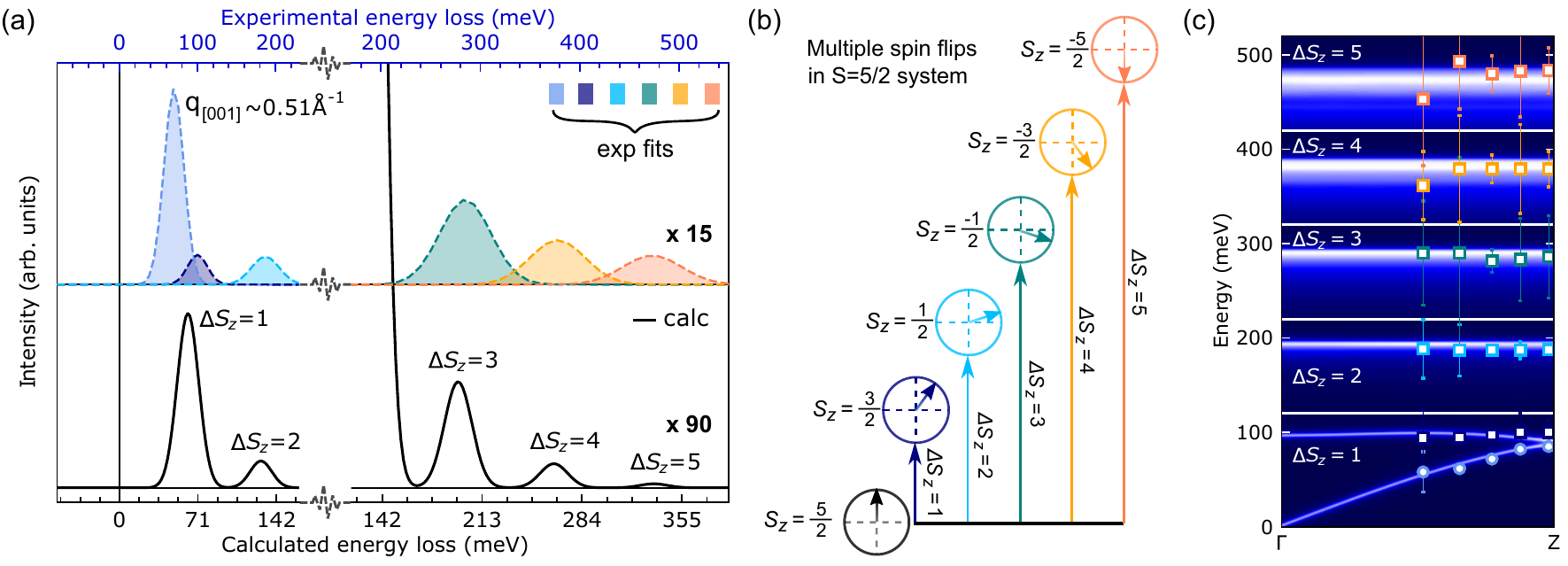}
            \caption{Calculations of magnetic excitations for $\alpha$-Fe$_2$O$_3$. (a) Calculated RIXS spectrum of multispin-flip excitations  using the LDA+DMFT AIM (black solid line). The color-filled Gaussians are the fitting results of the experimental RIXS data in Fig.~\ref{Fig2}(a). The elastic peak is removed in both spectra. The calculated spectral weight for $\Delta S_z=3,~4,~5$ underestimates the measured intensity, and possible arguments are presented in Sec.~VI of \cite{sm}. (b) A diagram of the possible multispin-flip excitations for an $S=\frac{5}{2}$ system, using a local picture. (c) The \textbf{\textit{q}}-resolved DOS of the AFM magnons. The single-magnon is calculated within the Heisenberg model. The multimagnons are obtained as a convolution of multiple single magnons, as explained in the text. For each $\Delta S_z$ mode, the DOS is normalized to the maximum achieved in the investigated \textbf{\textit{q}} range. The horizontal white lines define the energy windows used for each DOS normalization. The markers are the experimental RIXS energies identified in Fig.~\ref{Fig2}(c) for the acoustic (circles) and optical (squares) branches.
            }
            \label{Fig3}
        \end{figure*}
        
        To address the absolute energy and \textbf{\textit{q}}-dispersion of these excitations,
        we resort to the
        Heisenberg model $H=\sum_{i,j} J_{ij}{\bf S}_i \cdot {\bf S}_j$ with the exchange parameters of bulk $\alpha$-Fe$_2$O$_3$ ~\cite{Samuelsen69, Samuelsen70}. We derive first the dispersion $\varepsilon(\textbf{\textit{q}})$ of the single-magnon acoustic and optical branches, being the two eigenmodes of $H$ for $\alpha$-Fe$_2$O$_3$.
        For the higher-order spin-flip excitations ($\Delta S_z \geq 2$), we represent them as multiple, noninteracting magnons rather than bound states of several magnons.
        Therefore, a multimagnon excitation 
        at a momentum \textbf{\textit{q}} is expressed as a product (convolution) of 
        the single magnons with momenta (\textbf{\textit{q}}$_1$, \textbf{\textit{q}}$_2$, $\cdots$) constrained by the momentum conservation, i.e., \textbf{\textit{q}}=\textbf{\textit{q}}$_1 +$ \textbf{\textit{q}}$_2 +$ $\cdots$. 
        Evaluating the $n$-particle Green's functions $G_n(\omega, \textbf{\textit{q}})=[\omega-
        \varepsilon(\textbf{\textit{q}}_1)-\varepsilon(\textbf{\textit{q}}_2)-\cdots-\varepsilon(\textbf{\textit{q}}_n)]^{-1}$ for single magnons across the whole Brillouin zone, eventually produces the spectral density in energy-momentum space for the $\Delta S_z=n$ excitation.
        
        In Fig.~\ref{Fig3}(c), we display the resulting density of states (DOS) for $\Delta S_z=1,~2,~3,~4,$ and 5 multimagnons excitations.
        The $\Delta S_z=1$ acoustic and optical branches topping around 98~meV reproduce the experimental RIXS results (circle and square symbols overlaid to the DOS). 
        Although the calculated multimagnon spectra 
        present a broad continuum, the spectral weights actually accumulate in narrow peaks at multiples of approximately 98~meV. This reflects the flat nature of the optical magnon branch in a 3D antiferromagnet, contrary to a 1D system \cite{Nag22},
        and well captures the experimental RIXS results reproduced in Fig.~\ref{Fig3}(c).
        Thanks to this agreement, we fully corroborate the assignment of the measured harmoniclike peaks as $\Delta S_z=1,~2,~3,~4,$ and 5 magnetic excitations, and infer their character as multiple noninteracting magnons. The noninteracting behavior is an important aspect for the magnon propagation in  $\alpha$-Fe$_2$O$_3$, directly relevant for spintronics and consistent with the recently reported long-magnon-propagation length \cite{Lebrun18}.
        We further note that the total multimagnon contribution observed in $\alpha$-Fe$_2$O$_3$ amounts up to 10\% of the single-magnon spectral weight (including the effect of the RIXS cross section). 

        \section{Multimagnon excitations in RIXS}
        Ultimately, our observations raise the question of the origin of multimagnon excitations in the RIXS spectra,
        in light of the dipolar character of the photon emission and absorption processes exchanging at most two quanta of angular momentum. 
        To answer this, we investigate the spin character of the RIXS intermediate states $|m \rangle$. 
        To simplify the analysis, we employ an atomic model~\cite{Wang19, Groot2021}, which has the same symmetry properties as AIM, but allows explicit evaluation of the intermediate states $|m \rangle$ (see Sec.~VII of the Supplementary Material~\cite{sm} for details). We then examine the amplitudes 
        $
             {I_{m}=| \langle f| T_e |m\rangle \langle m | T_i |g\rangle |^2} \notag
        $ 
        [see Eq.~S1 in the Supplementary Material~\cite{sm}] for the transition from the ground state $|g\rangle$ to the final state $|f\rangle$ via $|m\rangle$ 
         \footnote{Note that in the LDA+DMFT AIM calculation a resolvent technique is used for computing the RIXS intensities, and thus such $|m\rangle$-resolved amplitudes ($I_m$) are not available~\cite{sm,Hariki20}}.
         \jl{Note that the amplitudes $I_{m}$ are only indicative of whether a given channel is available or not. They shall not be summed up to estimate the total RIXS amplitude, which arises due to interference of all the different channels.}
        
        Since the electric dipole transitions conserve the electron spin, the spin-flip excitations are enabled by spin-non-conserving terms in the system Hamiltonian. 
       It is therefore crucial to understand which are the active symmetry-breaking terms, and how they affect the final states in RIXS. 
       The initial state $|g\rangle$ and final states $|f\rangle$ of the studied RIXS process are orbital singlets, $S =\frac{5}{2}$ eigenstates of the spin $S_{3d}$, thanks to their half-filled 3$d^5$ character and the strong Hund's coupling dominating over the crystal field (CF). 
       These states ($|g\rangle$ and $|f\rangle$) remain approximately so even if the weak $3d$ spin-orbit-coupling SOC$_{3d}$ is considered. 
       The completely filled $2p$ core shell, instead, is subject to the strong spin-orbit-coupling SOC$_{2p}$ but negligible CF; thus, it can be labeled by
       the total angular momentum $J_{2p}$. 
       In the intermediate state $|i\rangle$, when a $2p$ electron is promoted to the $3d$ state, the core-valence interaction CV$_{2p \leftrightarrow 3d}$ \cite{Ghiasi19}
       becomes relevant as it entangles the partially filled $2p$ and $3d$ states due to its multipole terms. As a result, CV$_{2p \leftrightarrow 3d}$ conserves only the total angular momentum of both shells $J_{2p}+J_{3d}$, but not the $J_{2p}$ and $J_{3d}$ angular momenta separately. A summary of the conserved quantities for each active interaction is presented in Table~\ref{tab_main}.

        On the basis of this information, we disentangle the role of the different terms, i.e., SOC$_{2p}$, SOC$_{3d}$, CV$_{2p \leftrightarrow 3d}$, and CF, in the atomic model and uncover their contribution to $I_m$ by switching them on and off independently. 
        Considering the photon-electron interaction under dipole approximation, the following selection rule applies to the photon absorption or emission process, e.g., $\Delta J_z=0,~\pm1$ \cite{Ament11}. As we focus here on spin excitations, we need to transfer the available angular momentum to the spin of the electrons by activating proper spin-orbital-coupling channels. 
        
\begin{figure}[h!]
            \includegraphics[width=1.00\columnwidth]{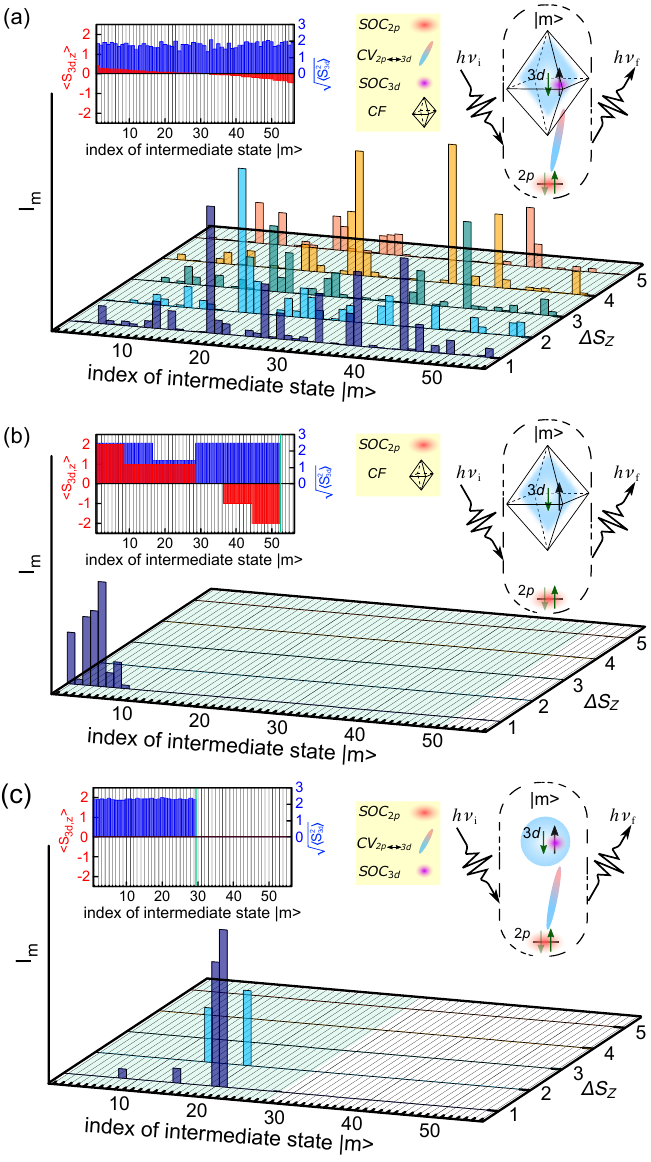}
            \caption{Calculated RIXS amplitude $I_{m}=| \langle f| T_e |m\rangle \langle m | T_i |g\rangle |^2$ of the magnetic excitations [\jl{the numerator part of the first term in the Kramers-Heisenberg formula}; see Eq.~S1 of the Supplementary Material~\cite{sm}] from the ground state $|g\rangle$ to the final state $|f\rangle$ via different intermediate states $|m \rangle$ with energies $E_m \in [E_{L_3} \pm 0.3~{\rm eV}]$ ($E_{L_3}$ denotes the Fe-$e_g$ absorption energy), using the local atomic model. (a) Amplitude $I_{m}$ when SOC$_{3d}$, SOC$_{2p}$, CV$_{2p \leftrightarrow 3d}$, and CF interactions are switched on. (b) Amplitude $I_{m}$ when SOC$_{3d}$ and CV$_{2p \leftrightarrow 3d}$ are off (i.e., SOC$_{2p}$ and CF are on). (c) Amplitude $I_{m}$ when CF is off (i.e., SOC$_{3d}$, SOC$_{2p}$, CV$_{2p \leftrightarrow 3d}$ are on).
            Within the selected $L_3$-edge energy window, we consider 56, 51, and 29 $|m \rangle$ states with different eigenenergies $E_m$ (indicated by the green shaded area), respectively, for (a), (b), and (c), accounting for 95\% of the total calculated RIXS intensity \textcolor{black}{(integrated over the energy loss) at a given photon energy \textcolor{black}{($E_{L_3}$)}}.
            The insets on the left of each plot summarize the $\langle\hat{S}_{3d,z}\rangle$ and $\sqrt{\langle\hat{\bf S}_{3d}^2\rangle}$ values on the Fe 3$d$ shell in these intermediate states.} 
            \label{Fig4}
        \end{figure}

        With all interactions on [see Fig.~\ref{Fig4}(a)], 
        a large number of nondegenerate $|m \rangle$ states is accessible, and their amplitude is finite for $\Delta S_z =1, 2, 3, 4$, and 5. With reference to Table~\ref{tab_main}, when these four interactions are simultaneously active, no quantity is conserved (a quantity is conserved when it is so for each active interaction). In particular, $S_{3d}$ is not a good quantum number for the intermediate states $|m \rangle$. This is reflected by  $\langle\hat{S}_{3d,z}\rangle$  and $\sqrt{\langle\hat{\bf S}_{3d}^2\rangle}$ differing from the integer $n$ and $\sqrt{n(n+1)}$, respectively; see inset of Fig.~\ref{Fig4}(a). 

        \begin{table}
            \centering
            \resizebox{\columnwidth}{!}{%
            \begin{tabular}{{l c c c c c c c}}
                \hline
                \hline
                 No. &    $J_{2p}+J_{3d}$  &  $J_{2p}$  &   $J_{3d}$    &   $L_{3d}$   &  $S_{2p}+S_{3d}$   & $S_{2p}$    & $S_{3d}$ \\
                \hline
                SOC$_{2p}$ & {\large \cmark} & {\large \cmark} & {\large \cmark} & {\large \cmark} & {\large \xmark} & {\large \xmark} & {\large \cmark} \\
                SOC$_{3d}$ & {\large \cmark} & {\large \cmark} & {\large \cmark} & {\large \xmark} & {\large \xmark} & {\large \cmark} &  {\large \xrmark}\\
                CV$_{2p \leftrightarrow 3d}$ & {\large \cmark} & {\large \xmark} & {\large \xmark} &  {\large \xmark} & {\large \cmark} & {\large \xmark} &  {\large \xrmark}  \\  
                CF &  {\large \xbmark} & {\large \cmark}& {\large \xmark} & {\large \xmark} &{\large \cmark}&{\large \cmark}&{\large \cmark}\\
                \hline
                \hline
            \end{tabular}
            }
            \caption{Conserved quantities for the $2p$ and $3d$ shells (columns), under SOC$_{2p}$, SOC$_{3d}$, CV$_{2p \leftrightarrow 3d}$, and CF interactions (rows). 
            \jl{The  {\large \cmark} marker indicates that the corresponding operators commute, while the  {\large \xmark} marker indicates  non-commuting operators.
            An observable is conserved if its column contains only {\large \cmark} markers, i.e., the interactions with {\large \xmark} mark are not included in the Hamiltonian.}
            The red color highlights the broken quantities that are critical for accessing multi-spin excitations, enabling the transfer of angular momentum to the electron spin. 
            The blue color highlights the broken quantity that expands the available angular momentum beyond $\Delta J_z=2$ exchanged by the absorbed and emitted photons. 
            }
            \label{tab_main}
        \end{table}

        In Fig.~\ref{Fig4}(b), we switch off the two terms that break $S_{3d}$, respectively, CV$_{2p \leftrightarrow 3d}$ and SOC$_{3d}$ (see Table~\ref{tab_main}). 
        In this case, $S_{3d}$ is conserved, as demonstrated by the integer value of, e.g.,  $\langle\hat{S}_{3d,z}\rangle$. A finite $I_m$ is obtained only for $\Delta S_z =1$ due to the spin mixing in the $2p$ state by SOC$_{2p}$ \cite{Braicovich10, Dean12, Pelliciari21_2};
        i.e., electrons with opposite spin can be excited or deexcited in the photon absorption or emission process.
        
        Comparing the two scenarios in Figs.~\ref{Fig4}(a) and 4(b), it clearly emerges the key role of CV$_{2p \leftrightarrow 3d}$ and SOC$_{3d}$ in allowing multispin excitations in the RIXS process. The nonconservation of $S_{3d}$ caused by the spin-orbital mix in the Fe $3d$ states enables the transfer of the angular momentum to the electron spin. This can happen directly by SOC$_{3d}$ or indirectly due to the combination of SOC$_{2p}$ and the multipole part of CV$_{2p \leftrightarrow 3d}$. In both cases, a number of states $|m \rangle$ have overlap with initial and final states of different $S_z$, leading to multispin-flip excitations.
        
        
        Finally, we discuss the role of CF. In Fig.~\ref{Fig4}(c), we keep all interactions on but we switch off the CF, i.e. in condition of spherical symmetry. The total angular momentum $J_{3d}+J_{2p}$ is conserved, but not $S_{3d}$. A finite $I_m$ is found for multispin excitations, but only up to $\Delta S_z=2$. In this case, in fact, only the angular momentum carried by the absorbed and emitted photons is the source of $\Delta S_z$, which is therefore less than or equal to 2. 
        On the contrary, when the spherical symmetry is broken, i.e. CF$\neq$0 as in Fig.~\ref{Fig4}(a), the $J_{3d}+J_{2p}$ quantity is not conserved and the crystal lattice acts as a reservoir of angular momentum, allowing the exchange of more than two quanta of $\Delta J_z$. Multispin excitations beyond $\Delta S_z$=2~\cite{Ghiringhelli09, Nag20} then become
        possible, reaching $\Delta S_z=2S$ in a spin-$S$ system, e.g. $\Delta S_z=5$ for $\alpha$-Fe$_2$O$_3$. \jl{Following this logic, we predict to resolve multi-spin excitations up to $\Delta S_z=3$ for CoO, an $S=\frac{3}{2}$ system.} 
        For a quantitative comparison of the individual strength of CV$_{2p \leftrightarrow 3d}$, SOC$_{3d}$, SOC$_{2p}$ onto the multispin-flip RIXS intensity, we refer to Fig.~S7 in the Supplementary Material~\cite{sm}.

        \section{Conclusion}
        In conclusion, we unveil magnons and their \textbf{\textit{q}}-dynamics in $S=\frac{5}{2}$ $\alpha$-Fe$_2$O$_3$ thin films, by combining RIXS measurements and calculations based on LDA+DMFT AIM and the Heisenberg model. Below 100~meV, we observe the single-magnon mode with equivalent energy in 15- and 30-nm films. This mode displays a {\textit{\textbf{q}}}-dispersion matching the one of the bulk, thus indicating the robustness of out-of-plane magnons in hematite down to 15~nm.
        Furthermore, an unexpected sequence of equally spaced
        excitations
        is observed
        in the ($100-500$)-meV range 
        assigned to
        $\Delta S_z=1, 2, 3, 4$ and 5 spin-flip transitions corresponding to multiple noninteracting magnons. Based on these results, we convey a comprehensive description of the energy, character, and momentum-dependence of magnons in hematite thin films, with potential impact on magnon transport (e.g., hematite offers noninteracting, long-range-travelling magnons with finite contribution from multimagnon modes) and on magnon pumping (e.g., pump frequency should be set away from the multimagnon energies to minimize dissipation channels). 
        At the same time, our study generalizes the use of $L$-edge RIXS as a multispin-excitation probe, being able to access magnetic excitations up to $\Delta S_z=2S$. We demonstrate that the breaking of $S_{3d}$, and the consequent spin-orbital mix in the valence shell, is the key element for opening the path to multispin-flip excitations. Furthermore, for $\Delta S_z > 2$, we introduce the key role of the lattice-breaking spherical symmetry acting as a reservoir of angular momentum and complementing the one carried by the absorbed and emitted photons. \jl{Such mechanism can generate interesting consequences. For example, in multiorbital systems (under proper conditions of e.g. resonant edge, crystal field symmetry, Hund's coupling), we could control the intensity of the spin and multispin-flip excitations by manipulating the local crystal environment through external perturbations, like e.g. strain, light pulses, gating, chemical pressure. This opportunity could represent a novel route to tune magnetic excitations in spintronic materials, deserving further studies.} 
        Ultimately,  our method proves the feasibility of magnon studies in device-like conditions, i.e., thin films, encouraging future investigations under, e.g., electric gating or photoexcitation.

        \textcolor{black}{
        As a final note, we underline that understanding multimagnons beyond $\Delta S_z>$~2 by means of a two-photon-scattering process is a very timely and active topic stimulated by the recently improved RIXS spectrometers. Indeed, similar conclusions on the key role of the CF in the context of multimagnons with $\Delta S_z>$~2 have also been reported by Elnaggar {\it{et al.}} while studying the azimuthal dependence of $\alpha$-Fe$_2$O$_3$ single crystals~\cite{Elnaggar22}.}

    \section{Acknowledgements}
    The authors thank M. P. M. Dean and T. Uozumi for fruitful discussions. This work is primarily supported by the U.S. Department of Energy (DOE), Office of Science, Basic Energy Sciences, Early Career Award Program (Brookhaven National Laboratory) and under Grant No. DE-SC0001304 (The Ohio State University). This research uses the beamline 2-ID of the National Synchrotron Light Source II, a DOE Office of Science User Facility operated for the DOE Office of Science by Brookhaven National Laboratory under Contract No. DE-SC0012704. A.H. is supported by JSPS KAKENHI Grants No. 21K13884 and No. 21H01003. The computations are performed at the Vienna Scientific Cluster.


\begin{thebibliography}{60}
\bibitem [{fn1()}]{fn1}%
  \BibitemOpen
  \href@noop {} {}\bibinfo {note} {In the LDA+DMFT self-consistent
  calculations, the double counting correction value $\mu_{\rm dc}$ is adjusted
  as a parameter. The charge-transfer energy $\Delta_{\rm CT}$ is defined as
  $\Delta_{\rm CT}=\varepsilon_d-\mu_{\rm dc}+5\times U_{\rm
  avr}-\varepsilon_p$, where $\varepsilon_d$ ($\varepsilon_p$) is the onsite
  energy of Fe 3$d$ (O 2$p$) states in the LDA result, and $U_{\rm avr}$ is the
  averaged Coulomb interaction on the Fe 3$d$
  shell~\cite{Higashi21,Winder20}.}\BibitemShut {Stop}%
\end{thebibliography}
\end{document}


    \title{Supplemental Material for: ``Single- and Multimagnon Dynamics in Antiferromagnetic $\alpha$-Fe$_2$O$_3$ Thin Films''}

    \author{Jiemin Li}
    \affiliation{National Synchrotron Light Source II, Brookhaven National Laboratory, Upton, NY 11973, USA.}
    \author{Yanhong Gu}
    \affiliation{National Synchrotron Light Source II, Brookhaven National Laboratory, Upton, NY 11973, USA.}
    \author{Yoshihiro Takahashi}
    \affiliation{Department of Physics and Electronics, Osaka Metropolitan University, 1-1 Gakuen-cho, Nakaku, Sakai, Osaka 599-8531, Japan.}
    \author{Keisuke Higashi}
    \affiliation{Department of Physics and Electronics, Osaka Metropolitan University, 1-1 Gakuen-cho, Nakaku, Sakai, Osaka 599-8531, Japan.}
    \author{Taehun Kim}
    \affiliation{National Synchrotron Light Source II, Brookhaven National Laboratory, Upton, NY 11973, USA.}
    \author{Yang Cheng}
    \affiliation{Department of Physics, The Ohio State University, Columbus, OH 43210, USA.}
    \author{Fengyuan Yang}
    \affiliation{Department of Physics, The Ohio State University, Columbus, OH 43210, USA.}
    \author{Jan Kune\v{s}}
    \affiliation{Institute of Solid State Physics, TU Wien, 1040 Vienna, Austria.}
    \author{Jonathan Pelliciari}
    \affiliation{National Synchrotron Light Source II, Brookhaven National Laboratory, Upton, NY 11973, USA.}
    \author{Atsushi Hariki}
    \affiliation{Department of Physics and Electronics, Osaka Metropolitan University, 1-1 Gakuen-cho, Nakaku, Sakai, Osaka 599-8531, Japan.}
    \author{Valentina Bisogni}
    \affiliation{National Synchrotron Light Source II, Brookhaven National Laboratory, Upton, NY 11973, USA.}

    \maketitle
    
    This supplemental material contains: I. sample information; II. energy dependence of the resonant inelastic x-ray scattering (RIXS) spectra across the Fe $L_3$-edge; III. thickness dependence of the low-energy excitations; IV. temperature dependence of the low-energy excitations; V. fitting of the \textbf{\textit{q}}-dependent RIXS data; VI. the LDA+DMFT Anderson impurity model (AIM) simulation of $\alpha$-Fe$_2$O$_3$ RIXS spectra; and VII. the atomic model simulation for multi-spin flip excitations.
    
    \section{Sample Information}
    Epitaxial $\alpha$-Fe$_2$O$_3$ films were grown on Al$_2$O$_3$ (001) substrate using off-axis sputtering at a substrate temperature of 500$^{\circ}$C. The films grew fully relaxed on top of Al$_2$O$_3$ (5\% of lattice mismatch), with the corundum structure $R3\Bar{c}$, typical of hematite. The lattice parameters defined within the hexagonal unit cell are $a$ = $b \sim$ 5.038\;\AA~and $c \sim$ 13.772\;\AA~\cite{Shull51}. Further details on the growth and characterization are collected in Ref.~\cite{Cheng20}. At room temperature, the spin moment of hematite lies in the (001) plane and stacks antiferromagnetically along the $c$--axis of the hexagonal cell. While a Morin transition around $T_M\sim~$261~K manifests in single crystals of $\alpha$-Fe$_2$O$_3$, owing to the temperature dependent axial magneto-crystalline anisotropy $K$(T)~\cite{Artman65}, this is essentially suppressed in thin films ~\cite{Fujii94, Gota01}, stabilizing the  planar spin arrangement also at low temperature.

    \begin{figure*}
        \centering
        {\includegraphics{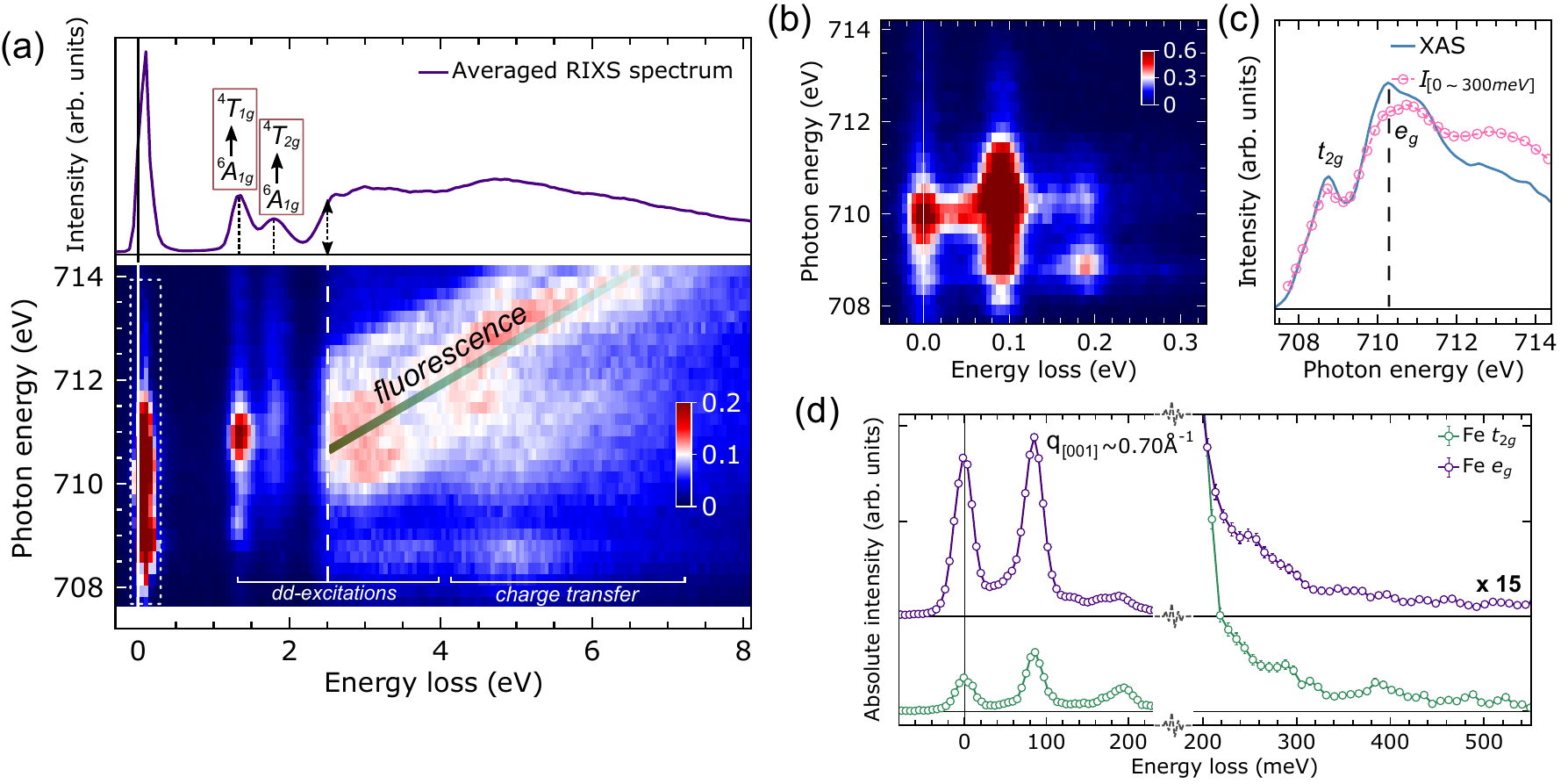}}
        \caption{The RIXS energy dependence across Fe $L_3$-edge at a geometry of $\theta=20^{\circ}$ and 2$\theta=150^{\circ}$ with linear-horizontally ($\pi$) polarized x-rays\jl{, and two representive RIXS spectra recorded at \textbf{\textit{q}$_{[001]}$ }$\sim$0.70$\AA^{-1}$ at Fe $t_{2g}$ and $e_g$ resonance}. (a) Overview of the RIXS energy map. The top spectrum is the averaged RIXS spectrum obtained from the ones displayed in the energy map. In there, the $dd$ excitations are indicated by dashed lines and the onset of the fluorescence is marked by a double-arrow. (b) Zoomed view of the low-energy excitations indicated by the white, dashed box in (a). \jl{(c)}  The XAS collected through partial fluorescence yield mode with the RIXS spectrometer (blue line), together with the RIXS intensity integral $I$ computed in the 0$\sim$300~meV energy range (pink dotted line). The dashed line indicates the incident photon energy used for RIXS measurements. The temperature was set to 100~K for the energy map. \jl{(d) RIXS spectra collected at \textbf{\textit{q}$_{[001]}$ }$\sim$0.70$\AA^{-1}$ ($\theta$=70$^{\circ}$ and 2$\theta$ =150$^{\circ}$) with photon energy tuned to Fe $t_{2g}$ and Fe $e_g$ peaks of XAS.}}
        \label{Fig_sm1}
    \end{figure*} 
    
    \section{Energy dependence}
    
    Figure~\ref{Fig_sm1} (a) displays an overall picture of the energy dependence of the RIXS spectra across the Fe $L_3$-edge, together with the averaged RIXS spectrum (see top plot), generated by integrating the RIXS spectra at all incident photon energies. The high energy features (above 1 eV) are discussed in main text. Here we focus on the low-energy excitations $<$~500~meV:  intense features appear in the neighborhood of the elastic line, see the strong peak centered around $\sim100$~meV in the averaged RIXS spectrum of Fig.~\ref{Fig_sm1}(a). To better visualize these excitations, we display a zoomed-view in Fig.~\ref{Fig_sm1}(b). Several strong inelastic peaks below 300~meV clearly stand out in the color map, well separated from the elastic peak, thanks to the high energy resolution of $\Delta E$=23~meV used for these measurements. These excitations are found to resonate at the incident energy of 710.2~eV (e$_g$ resonance), \jl{see the RIXS integrated intensity in the 0--300~meV energy range (pink dotted line) displayed in Fig.~\ref{Fig_sm1}(c)}. The Fe e$_g$ resonance is then used for all the RIXS measurements displayed in this work. \jl{In Fig.~\ref{Fig_sm1}(d), we display two representive RIXS spectra recorded at both Fe $t_{2g}$ and $e_g$ resonance respectively. Although the low-energy excitations are overall weaker at Fe $t_{2g}$ resonance, the scaling factor is however not constant for different excitations}.

    \begin{figure}
        \centering
        {\includegraphics[width=0.85\columnwidth]{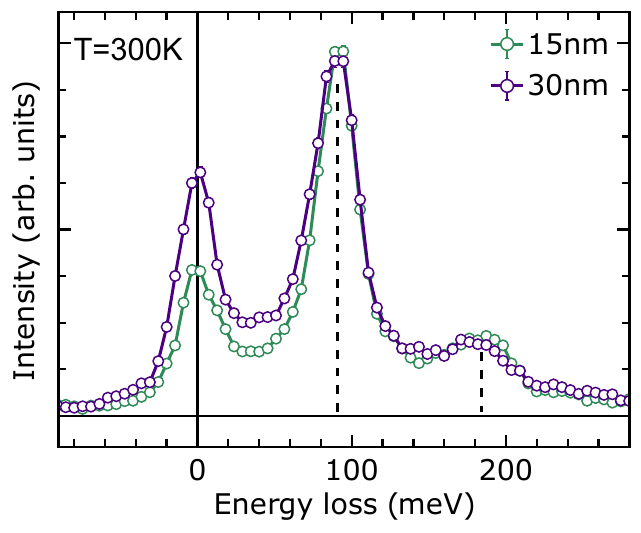}}
        \caption{The low-energy RIXS excitations of $\alpha$-Fe$_2$O$_3$ thin films with different thickness, 15~nm and 30~nm, recorded at a geometry of $\theta=20^{\circ}$ and 2$\theta=150^{\circ}$ with circular polarized x-rays. The dashed lines indicate the single-magnon ($\Delta S_z=1$) and the two-magnon ($\Delta S_z=2$) peaks. Both spectra were normalized to the integrated intensity over the whole energy loss range. The temperature was set to 300~K.}
        \label{Fig_sm2}
    \end{figure} 
        
    \section{Thickness dependence}
    Fig.~\ref{Fig_sm2} shows RIXS spectra measured on two thin films of $\alpha$-Fe$_2$O$_3$ with a thickness of 15~nm (green line) and 30~nm (blue line). The x-ray photon energy was set to the $e_g$ resonance. Circular-polarized light was used in this case to enhance the incoming photon flux, helpful for measuring the the thinner sample. The temperature was set to 300 K. 
    
    In order to directly compare the low-energy RIXS excitations as a function of thickness, we normalized the two spectra to the overall integrated area. Interestingly, the $\sim~100$~meV peak and the $\sim~200$~meV peak, interpreted in the text as $\Delta S_z=1$ and $\Delta S_z=2$ magnons respectively, appear at same energy positions in two samples and with comparable intensity. These findings imply that the magnon modes in hematite films are extremely robust down to 15~nm, and they are not subject to renormalization caused by electronic reconstruction or confinement \cite{Pelliciari21_2, Pelliciari21}. Finally, we note that the elastic line intensity of the two RIXS spectra displays a different spectral weight versus thickness, being stronger for the 30~nm film. This is likely due to a rougher surface for the thick film.
    
    \begin{figure}
        \centering
        {\includegraphics[width=1\columnwidth]{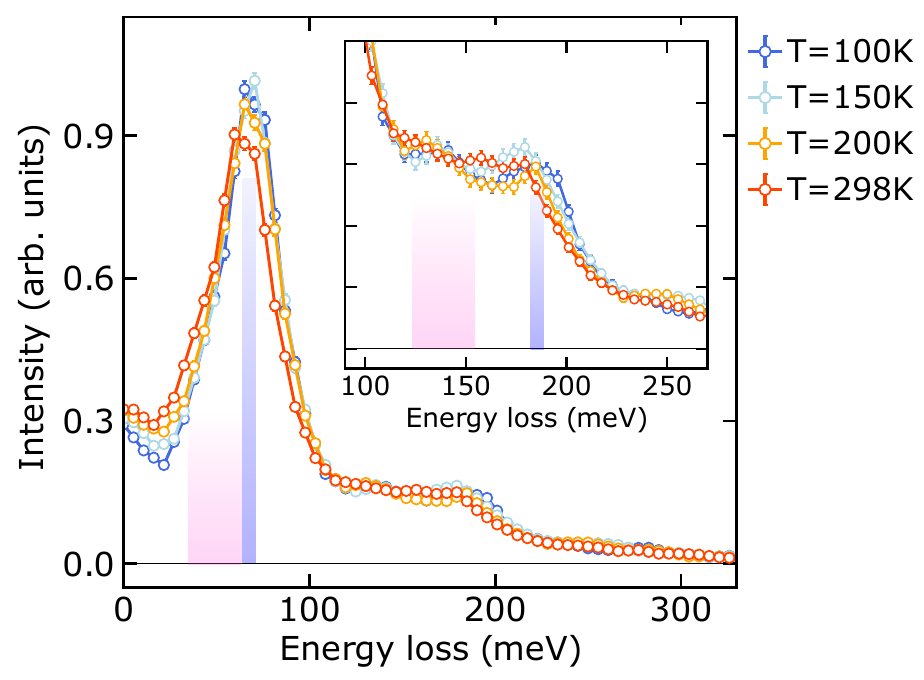}}
        \caption{Temperature dependence (from 100~K to 298~K) of the low--energy excitations at \textbf{\textit{q}} $\sim0.51$\;\AA$^{-1}$. The blue shaded area defines magnetic excitations, while the pink shaded area identifies phonons. The inset is a zoomed-in view on excitations around 200~meV.}
        \label{Fig_sm3}
    \end{figure}

    \section{Temperature dependence of low energy excitations}
    The thermal evolution is usually different for excitations involving different electronic degrees of freedom, such as the lattice and spin. This makes the temperature dependence of the RIXS spectra a suitable method for clarifying the nature of the observed excitations. Fig.~\ref{Fig_sm3} shows the temperature dependence of low energy excitations up to 300~meV at \textbf{\textit{q}} $\sim0.51$\;\AA$^{-1}$. The spectral weight between 30--70~meV indicated by the pink shaded area becomes slightly stronger and broader at higher temperature, a classic characteristic of phonons due to increasing thermal vibrations as captured by RIXS~\cite{Pattanayak19, Ament11}. The main peak at $\sim$ 80~meV, indicated by the blue shaded area, manifests instead a red-shift when warming up the sample and it becomes slightly weaker in intensity. This behavior is consistent with magnetic excitation, i.e. single-magnon ($\Delta S_z=1$) in this case as reported in previous hematite studies \cite{Samuelsen70}. This provides an experimental evidence for the magnetic nature of the $\sim$100~meV peak. In the inset of Fig.~\ref{Fig_sm3}, we show the details of the spectral weight in the 100 - 200~meV range. The peak around 200~meV (indicated by the blue shaded area) displays a similar red-shift behavior as the single-magnon, suggesting a magnetic origin for this peak as well, interpreted in the text as a multi-magnon corresponding to $\Delta S_z=2$. The nature of this peak was heavily debated before and then proved being magnetic excitation by the Raman scattering through the effect of pressure and isotope substitution \cite{Massey90}. Furthermore, the additional spectral weight present around 140~meV (indicated by the pink shaded area) gets slightly broader with temperature. The phonon structure of $\alpha$-Fe$_2$O$_3$ obtained by infrared reflection spectroscopy, Raman scattering, and lattice dynamics studies reveals several phonon branches below $\sim$70~meV (as identified by the pink area in Fig.~\ref{Fig_sm3} centered around 40 meV and partially overlapping with the single-magnon in our data), two-phonon mode around 145~meV and potentially three-phonon mode around 225~meV \cite{Chamritski05, Pattanayak19,Onari77,Beattie70,Kappus75,Azuma05}. Therefore, we ascribe the spectral weight around $\sim$~140~meV to the two-phonon mode. Since a finite but weak spectral weight is visible in the RIXS spectra around 225~meV, it is reasonable to ascribe this intensity to the three-phonon mode, although we cannot extract any valuable temperature dependence for it.

    \section{\textbf{\textit{q}}-dependent data and their fitting}
    The momentum dependence of the RIXS excitations was investigated along the [001]--direction (film out-of-plane) by changing the scattering angle (2$\theta$) with the sample $\theta$--angle approximately at half of the 2$\theta$--angle. The reached \textbf{\textit{q}$_{[001]}$} values are listed in the Table~\ref{T1} for each pair of $\theta$--2$\theta$.  More specifically, the $\theta$--angle was set slightly off the exact specular reflection ($\leq 5 ^\circ$) to reduce the elastic peak. This causes a finite projection of photon momentum transfer perpendicular to the [001]--direction, i.e. \textbf{\textit{q}$_{[120]}$} as listed in Tab.~\ref{T1}. Since \textbf{\textit{q}$_{[120]}$} assumes reasonalby small value of less than 10\% \textbf{\textit{q}$_{[001]}$}, we are confident that the RIXS spectra are fully representative of the excitation dispersion along the [001]--direction. 
    
    \begin{table}[b]
        \begin{ruledtabular}
        \begin{tabular}{c c c c}
             \textbf{\textit{q}}$_{[001]}$ (\AA$^{-1}$)  &   \textbf{\textit{q}}$_{[120]}$ (\AA$^{-1}$)  &   $\theta$ ($^{\circ}$)  &   2$\theta$ ($^{\circ}$) \\
            \hline
            0.32    &  0          & 26.5      & 53      \\
            0.41    &  0.04          & 30      & 70      \\
            0.51           &  0          & 45      & 90       \\
            0.65            &  0.06          & 60      & 130       \\
            0.70            &  0.06       & 70     & 150       \\

        \end{tabular}
        \label{Tab1}
        \end{ruledtabular}
        \caption{The \textbf{\textit{q}} projections $vs$ $\theta$--$2\theta$ angles used for the RIXS experiment.}
        \label{T1}
    \end{table}

    \begin{figure*}
        \centering
        {\includegraphics{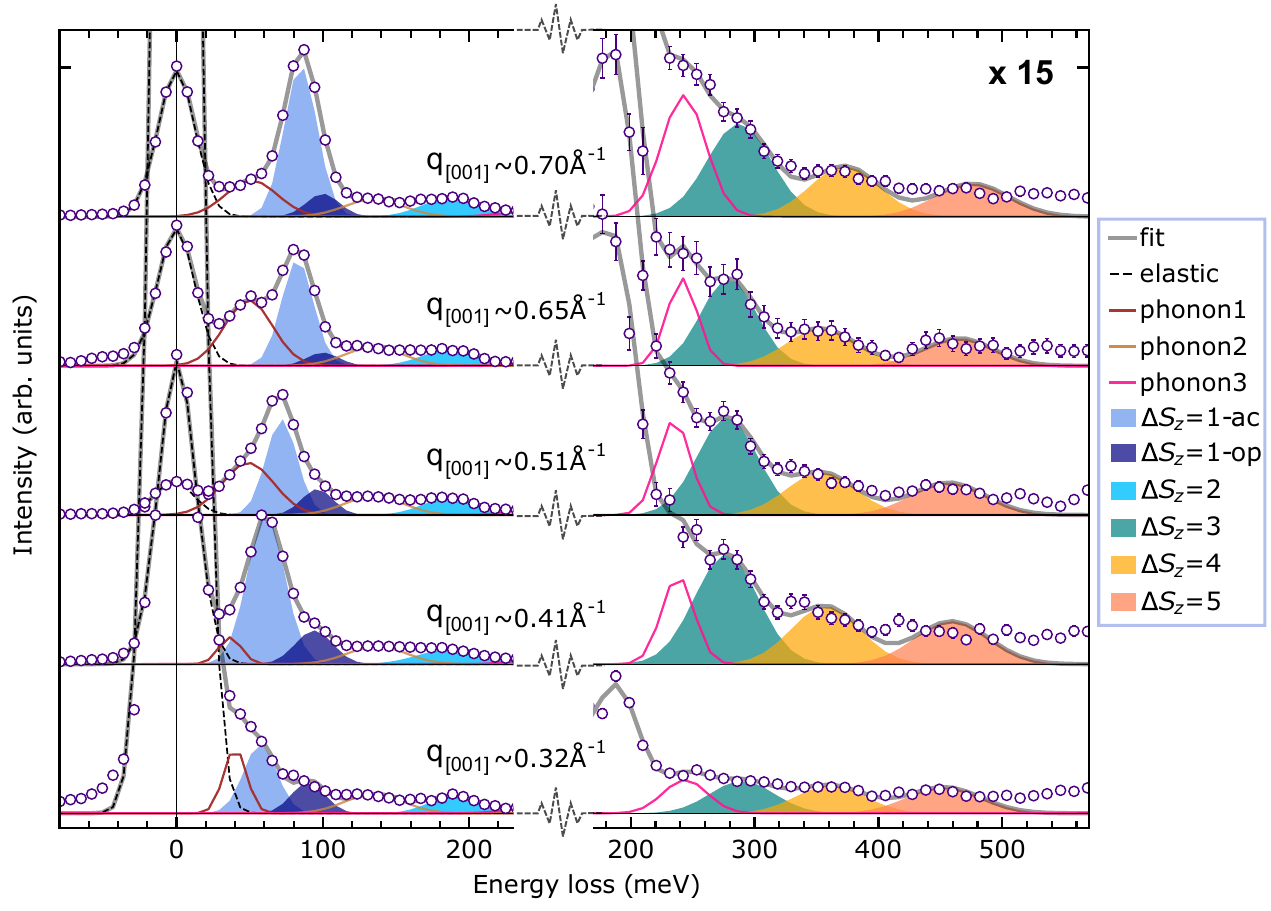}}
        \caption{The momentum dependence of low--energy excitations excited at the Fe $3d$--$e_{g}$ incident energy. The spectra were normalized to the integral of the high energy spectral weight (0.6~eV $\sim$ 8~eV).}
        \label{Fig_sm4}
    \end{figure*}

    Figure~\ref{Fig_sm4} displays the RIXS spectra recorded at five different \textbf{\textit{q}} positions. Due to the weak RIXS intensity beyond $\sim$200~meV, the data in the 200 -- 500 meV range were scaled by a factor of 15. As a function of momentum, the peak below 100~meV shifts to higher energy at larger \textbf{\textit{q}}, while all other peaks seem to remain at constant energies. To quantitatively extract the peak position for these low-energy excitations, we \textcolor{black}{fitted the RIXS spectra in the [-100, 570] meV range} 
    with ten contributions as listed in the following Equation:
            
    \begin{widetext}
        $I_{RIXS} = I_{elastic}+I_{phonon1}+I_{\Delta S_z=1_{ac}}+I_{\Delta S_z=1_{op}}+I_{phonon2}+I_{\Delta S_z=2}+I_{phonon3}+I_{\Delta S_z=3}+I_{\Delta S_z=4}+I_{\Delta S_z=5}$,
        \label{eqn:fit}
    \end{widetext}
    where the $I_{elastic}$, $I_{\Delta S_z=1_{ac}}$ and $I_{\Delta S_z=1_{op}}$ are constrained in width by the instrument energy resolution ($\Delta E \sim 23$~meV), while all other magnetic excitation widths were left free. In this model, we also included three Gaussians ($I_{phonon1}$, $I_{phonon2}$ and $I_{phonon3}$) with the centers around 40~meV, 145~meV and 225~meV to account for the low-energy phonons, two-phonon and three-phonon contributions as discussed in Sec.~IV. The widths of $I_{phonon1}$, $I_{phonon2}$ and $I_{phonon3}$ \textcolor{black}{were allowed to be $\geq \Delta E$. The energies for the magnetic excitations, $I_{\Delta S_z=1...5}$, were left free during the fitting. Being the resulting splitting between each pair of multi-magnons $\sim 100$ meV $>> \Delta E$, this minimizes the cross-talk between the fitted peaks, supporting the selected fitted model. At \textbf{\textit{q}} = 0.41$\AA^{-1}$ and 0.32$\AA^{-1}$ (where the overall RIXS cross-section weakens at low $2\theta$ angles), the weaker signal detected for $\Delta S_z$=4, and 5 challenged the fit. Therefore, only for these two spectra we constrained the distance between the $\Delta S_z$=4, and 5 peaks being larger than $\Delta E$, to avoid overlap.} 
    
    The resulting multi-magnon peak positions are summarized in Fig. 2(c) of the main text. \textcolor{black}{The good reproducibility obtained for each multi-magnon energy as a function of \textbf{\textit{q}} confirms the overall robustness of the implemented approach.} 
    Interestingly, except the mode below 100~meV showing a strong dispersion, all others do not disperse and appear at energies following an harmonic sequence with respect to the first one at $E_0$, i.e., $E_0$ ($\sim$~97$\pm$1 meV),  2$E_0$ ($\sim$~188$\pm$4 meV), 3$E_0$ ($\sim$~286$\pm$15 meV), 4$E_0$ ($\sim$~376$\pm$17 meV) and 5$E_0$ ($\sim$~478$\pm$19 meV), see the discussion in main text.
    
    \begin{figure}
        \centering
        {\includegraphics[width=0.85\columnwidth]{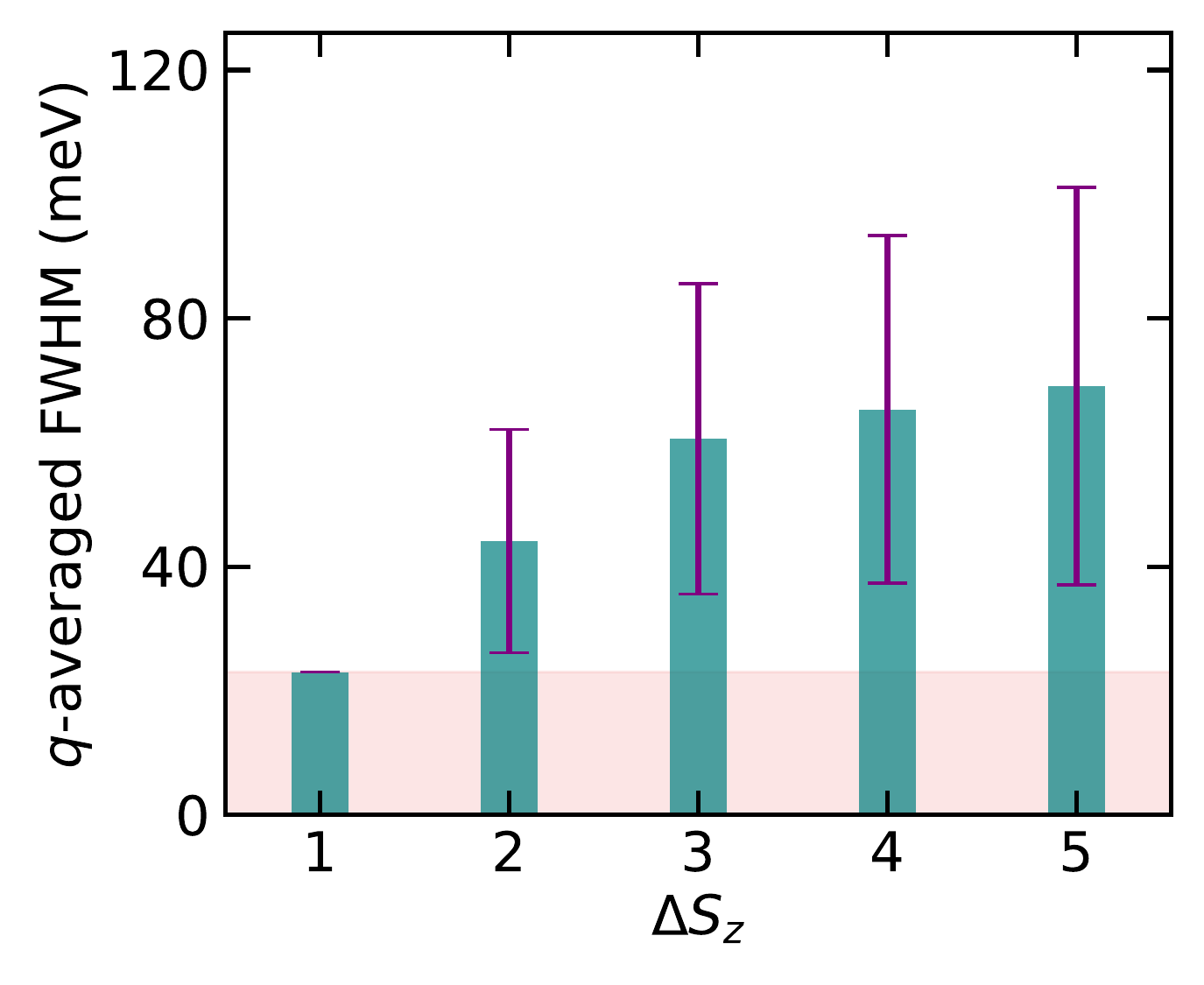}}
        \caption{\jl{The \textbf{\textit{q}}-averaged full width at half maximum (FWHM) of multi-spin-flip excitations extracted from the fitting analysis. The purple is the fitting error bars for each spin excitation. The shaded pink area indicates the experimental energy resolution (23meV).}}
        \label{Fig_fwhm}
    \end{figure}
    
    \jl{Finally, we comment on the width of the multi-spin-flip excitation for $\Delta S_z \geq 2$.  In Fig.~\ref{Fig_fwhm} we display the \textbf{\textit{q}}-averaged widths extracted from the fitting analysis. While the width of $\Delta S_z = 1$ excitation is constrained to the energy resolution, the width of the high-order excitations ($\Delta S_z \geq 2$) increases monotonically and saturates at $\sim$70 meV for $\Delta S_z = 5$. As explained in the main text, a higher spin-flip excitation does not represent a bound state (i.e. a mode), but several free (single spin-flip) magnons. Thus, the spectrum of such excitations is not a dispersion curve, with a width reflecting the excitation lifetime; rather, it is a continuum consisting of multiple magnon combinations simply respecting the energy and the momentum conservation.}



\section{LDA+DMFT modeling of $\alpha$-F\lowercase{e}$_2$O$_3$ and RIXS simulation}
    
    \begin{figure*}[h]
        \centering
        {\includegraphics[width=2\columnwidth]{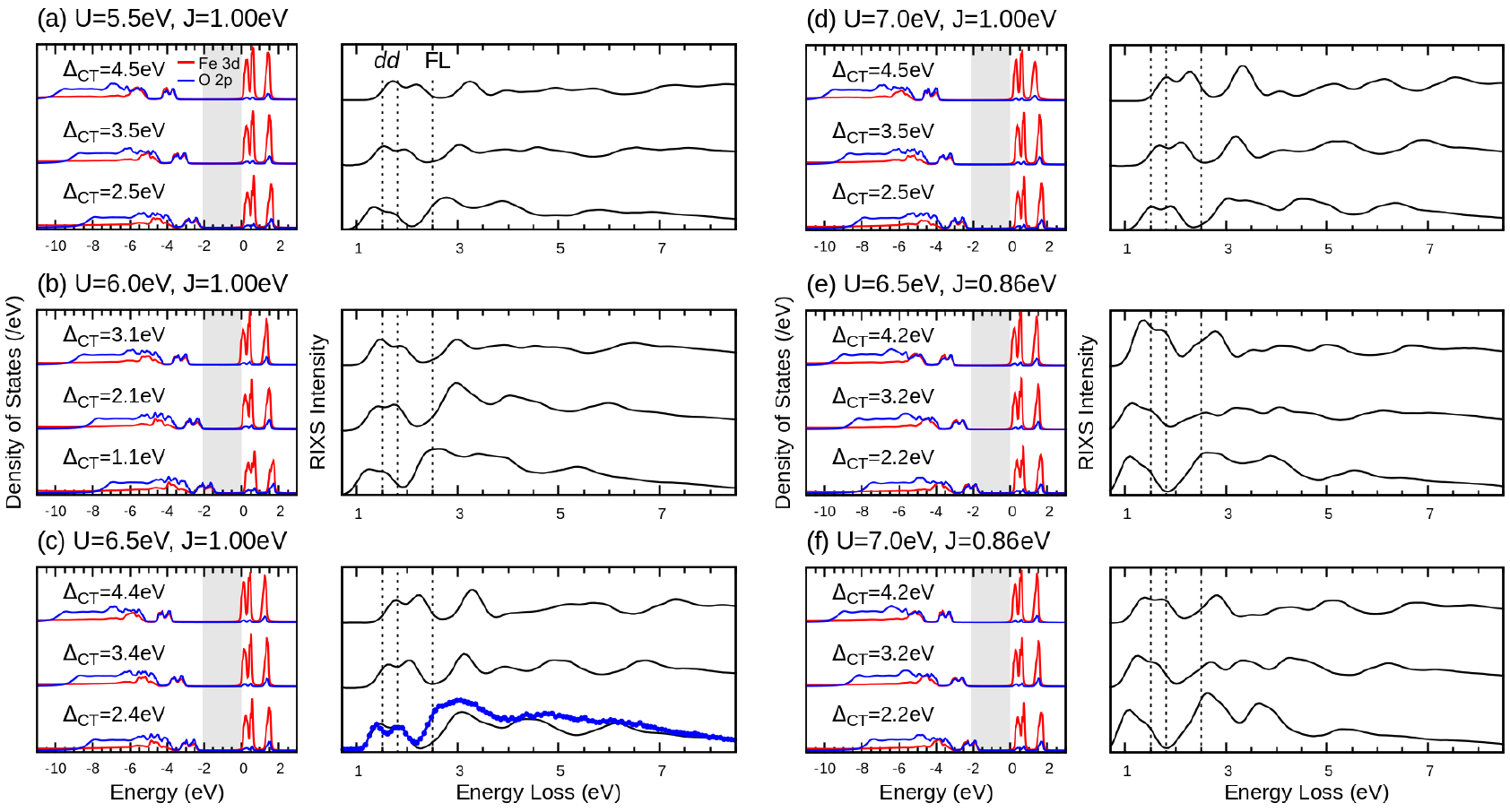}}
        \caption{The LDA+DMFT density of states (left) and the RIXS intensities (right) for selected model parameter values ($U$, $J$, $\Delta_{\rm CT}$).      
        The blue dotted line in panel (c) shows the experimental RIXS data presented in the main text. The parameters used for panel (c) provide the best agreement with the experiment, as highlighted by the three vertical dashed lines marking the $dd$ excitations and the onset of the charge excitation.}
        \label{Fig_sm5}
    \end{figure*}     
    
The LDA+DMFT simulation for the Fe $L_3$-edge RIXS spectra of $\alpha$-Fe$_2$O$_3$ proceeds in the two steps. Firstly, a standard LDA+DMFT calculation is performed for the experimental crystal structure of $\alpha$-Fe$_2$O$_3$~\cite{Finger80}.
Then the RIXS intensities are computed from the Anderson impurity model (AIM) with the LDA+DMFT-optimized hybridization densities $V^2(\varepsilon)$. In the first step, the LDA bands obtained with the WIEN2K package~\cite{wien2k} are projected onto a tight-binding model spanning the Fe 3$d$ and O 2$p$ bands~\cite{wien2wannier,wannier90}. The tight-binding model is then augmented by the electron-electron interaction within the Fe 3$d$ shell. The interaction is parameterized by Hubbard $U=F_0$ and Hund's $J=(F_2+F_4)/14$ where $F^0$, $F^2$, and $F^4$ are the Slater integrals~\cite{pavarini1,pavarini2}. 
To account for the electron-electron interaction already presented in the LDA description, the Fe 3$d$ orbital energy is shifted from its LDA value $\varepsilon^{\rm LDA}_d$ by the double-counting correction $\mu_{\rm dc}$~\cite{Karolak10,Kotliar06}. 
Practically, the $\mu_{\rm dc}$ renormalizes the splitting of Fe 3$d$ and O 2$p$ states, and thus relates to the charge-transfer (CT) energy $\Delta_{\rm CT}$. Imitating $\Delta_{\rm CT}$ used in the cluster model analysis of core-level spectra, we define the CT energy as $\Delta_{\rm CT}=\varepsilon^{\rm LDA}_d-\mu_{\rm dc}+5U_{dd}-\varepsilon^{\rm LDA}_p$~\cite{Hariki20,Higashi21}. We determine the three model parameters ($U$, $J$, $\Delta_{\rm CT}$) by comparing the calculated spectra with the present high-resolution Fe $L_3$ RIXS experiment as well as the reported value of the band gap ($E_{\rm gap}=2.1$~eV)~\cite{Kuhaili12}. The strong-coupling continuous-time quantum Monte  Carlo impurity solver~\cite{Werner06,Boehnke11,Hafermann12,Hariki15} was used to compute the local self-energy from the AIM in the self-consistent DMFT calculation. After reaching the self-consistency, the hybridization density $V^2(\varepsilon)$ is computed on the real energy axis with the analytically-continued self-energy obtained by the maximum entropy method~\cite{Jarrell96}. 
In the LDA+DMFT calculation, the antiferromagnetic (AF) magnetic order develops by breaking the spin rotational symmetry in the self-consistent calculation, which is manifested as the spin-dependence in self-energy computed from the AIM. 
We allow the spin-dependence explicitly in the LDA+DMFT calculations 
for the experimental magnetic unit cell, and obtained a stable AF solution for all the studied parameter values.
As a result, the hybridization density $V^2(\varepsilon)$ of AIM, representing the amplitude of the electron exchange between the impurity Fe site and the rest of the crystal~\cite{Ghiasi19}, depends on the spin. 
In the LDA+DMFT self-consistent calculation, the small spin-orbit coupling (SOC) on the Fe 3$d$ shell is not included. In the RIXS calculation described below, then the SOC is included to the AIM Hamiltonian $\hat{H}_{\rm AIM}$ so that the quantization axis ($z$-axis)  for the spin space lies on the [120] axis for the central Fe site in the left panel of Fig.~1(a) in the main text. The Fe spin value is discussed below.

The Fe $L_3$-edge RIXS spectra are computed from the AIM with the DMFT hybridization densities $V^2(\varepsilon)$, where the Fe 2$p$ core orbitals and its interaction with the Fe 3$d$ electrons are included explicitly. We treat the full-multiplet interaction in the core-valence (CV$_{2p \leftrightarrow 3d}$) interaction and the SOC in the core (2$p$) and valence (3$d$) orbitals, as described in Ref.~\cite{Hariki20}. 
The SOC in the 2$p$ (SOC$_{2p}$) and 3$d$ (SOC$_{3d}$) shell, and the Slater integrals defining the multipole part of the CV$_{2p \leftrightarrow 3d}$ interaction ($F_2$, $G_1$, $G_3$) are calculated with an atomic Hartree-Fock code. The actual values used in the simulation are summarized in Table ~\ref{t_soc}.
The $F_2$, $G_1$, and $G_3$ values are scaled down to 80\% of the Hartree-Fock values to simulate the effect of intra-atomic configuration interaction from higher basis configurations, which is a successful empirical treatment for $L$-edge spectroscopies of 3$d$ transition metal compounds~\cite{Matsubara05,Sugar72,Tanaka92,Groot90,Cowan18,Ghiasi19}.

We employ the configuration-interaction (CI) solver to compute the RIXS intensities directly on real frequencies following Kramers-Heisenberg formula, 
\begin{align}
F_{\rm RIXS}(\omega_{\rm out},\omega_{\rm in})&=\sum_{f} \left| \sum_{m}
     \frac{\langle f | T_{\rm e} | m\rangle \langle m | T_{\rm i} | g \rangle }
     {\omega_{\rm in}+E_g-E_m+i\Gamma}
      \right|^2 \notag \\ 
   &\times \delta(\omega_{\rm in}+E_g-\omega_{\rm out}-E_f) \\
   =&\sum_{f} \left| 
     \langle f | T_{\rm e}
     \frac{1}{\omega_{\rm in}+E_g-\hat{H}_{\rm AIM}+i\Gamma}  T_{\rm i} | g \rangle
      \right|^2 \notag \\
     &\times  \delta(\omega_{\rm in}+E_g-\omega_{\rm out}-E_f). 
     \label{eq:rixs}
\end{align}
Here, $|g\rangle$, $|m\rangle$, and $|f\rangle$ represent the initial, intermediate and final states with energies $E_m$ and $E_f$, respectively.
$\Gamma$ is the inverse lifetime of the core hole in the intermediate state, and is set to 300~meV~\cite{Hariki20}.
In practice, $V(\varepsilon)$ is presented by discretized levels (30 levels for each orbital and spin in the CI solver).
In the numerical evaluation, we employ the Lanczos method and the conjugated gradient method~\cite{Hariki20,Hariki17}. The explicit form of the AIM Hamiltonian $\hat{H}_{\rm AIM}$ is given in Ref.~\cite{Hariki20}.

Obtaining the spectrum of the eigenstates $\{|m\rangle\}$ in Eq.~(S1) is a hopeless task computationally for the AIM Hamiltonian $\hat{H}_{\rm AIM}$ with a large dimension.
The conjugated gradient method allows us to compute the propagation of the system with a 2$p$ core hole in Eq.~(S2) efficiently~\cite{Hariki20}.
To compare the computed RIXS spectra with the experimental data, a spectral broadening is taken into account using a Gaussian 
of 23~meV (full width at half maximum), simulating the experimental resolution for the low-energy RIXS features shown in Fig.~2(a) of the main text. To simulate the high-energy RIXS features with a larger lifetime broadening shown in Fig.~\ref{Fig_sm5} (c), we considered a Lorentzian of 200~meV which well reproduces the experimental line width.

        \begin{table}
            \begin{ruledtabular}
            \begin{tabular}{ c | l | l | c c c }
                  Config.  & SOC$_{3d}$\quad\quad & SOC$_{2p}$\quad\quad &  & CV$_{2p \leftrightarrow 3d}$ &  \\
                \hline\\[-10pt]  
                   & & & $F_{2}$ & $G_{1}$ & $G_{3}$  \\
                \hline\\[-10pt]  
                $d^{5}$ & 0.059  & & & &  \\
                $d^{6}$ & 0.052  & & & &  \\
                $d^{7}$ & 0.046  & & & &  \\
                \hline\\[-10pt]
                $\underline{p} d^{6}$ & 0.074 & 8.199 & 5.957 & 4.453 & 2.533 \\
                $\underline{p} d^{7}$ & 0.067 & 8.200 & 5.434 & 4.003 & 2.275 \\
                $\underline{p} d^{8}$ & 0.059 & 8.202 & 4.914 & 3.574 & 2.030 \\
            \end{tabular}
            \label{Tab_sk}
            \end{ruledtabular}
            \caption{The SOC$_{3d}$, SOC$_{2p}$, and Slater integrals for the core-valence multiplet interaction (CV$_{2p \leftrightarrow 3d}$) in the AIM Hamiltonian used in the Fe $L$-edge RIXS simulation. Here, $\underline{p}$ denotes a core hole in the Fe 2$p$ shell present in the intermediate states of the RIXS process. The values are evaluated using the atomic Hartree-Fock calculation, see text for details. All values are in eV.}
            \label{t_soc}
        \end{table}

Previous density function theory (DFT)-based and/or spectroscopy studies estimated $U$= $6 \sim 7$~eV and Hund's $J$ = $0.8\sim 1.0$~eV for Fe$_2$O$_3$~\cite{Kunes09b,Anisimov91,Greenberg18,Fujii99,Leonov19,Miedema15}. We thus started the calculation of the valence density of states (DOS) and RIXS spectra around these values. As aforementioned, the $\Delta_{\rm CT}$ modulates the splitting between Fe 3$d$ and O 2$p$ states, therefore related to the band gap ($E_{\rm gap}=2.1$~eV) of $\alpha$-Fe$_2$O$_3$ which is the energy cost to create unbound electron-hole pairs that determine the onset of the fluorescence features ($\sim$~2.5~eV, see the double-arrow in Fig.~\ref{Fig_sm1}) in RIXS spectra. Via the chemical bonding and intra-atomic multiplet interactions, the $U$ and $J$ together with the $\Delta_{\rm CT}$ strongly affect the energies of the first two $dd$ excitations ($\sim$1.4 and 1.9~eV from the experimental data) that correspond to $^6A_{1g}\rightarrow^4T_{1g}$ and $^6A_{1g}\rightarrow^4T_{2g}$ transitions flipping both spin ($S=\frac{5}{2} \rightarrow \frac{3}{2}$) and orbital ($e_g \rightarrow t_{2g}$) respectively. Fig.~\ref{Fig_sm5} summarizes the comparison of high energy features between the measured RIXS spectra and calculations computed from different parameter sets. We found that the parameter set $(U, J, \Delta_{\rm CT})=(6.5, 1.0, 2.4)$ (in eV unit) well reproduces the characteristic RIXS features (two $dd$ excitations and the onset of the fluorescence-like charge excitation) as well as the experimental band gap, see Fig.~\ref{Fig_sm5}(c). 

        \begin{table}
            \begin{ruledtabular}
            \begin{tabular}{c | c c c c c c c }
                 $ E_{\rm loss}$ (meV)  & 0 & 66 & 132 & 200 & 269 & 338   \\
                 \hline\\[-7pt]
                $ \langle \hat{S}_{z} \rangle $ &  2.27 &  1.35 &  0.43 & -0.48 & -1.40 & -2.30   \\
                $\Delta S_{z}$  & 0 & 0.92 & 1.84 & 2.75 & 3.67 & 4.57  \\
            \end{tabular}
            \label{Tab2}
            \end{ruledtabular}
            \caption{The excitation energy ($E_{\rm loss}$) and the expectation value of the spin operator $\hat{S}_z$ for the low-energy states in the LDA+DMFT AIM with the optimized parameter set.}
            \label{t_spin}
        \end{table}
        
Equipped with these optimized parameters, we examined the low-energy features of calculated RIXS spectra. As shown in Fig.~3(a) of main text, five inelastic peaks appear at $E_{\rm sf}$, 2$E_{\rm sf}$, 3$E_{\rm sf}$, 4$E_{\rm sf}$, and 5$E_{\rm sf}$ with $E_{\rm sf}=66$~meV and with decreasing intensities. 
To dig out the nature of these peaks, we inspected the expectation value of the spin operator $\hat{S}_z$ corresponding to these inelastic features, see Table~\ref{t_spin}. 
The values unambiguously validate that these five inelastic peaks are local (single-site) multi-spin-flip excitations with $\Delta S_z\approx 1, 2, 3, 4, 5$ to the AF ground state.

As described in the main text, the LDA+DMFT AIM describes only local spin-flip excitations against the source field generated by the bath (environment).
Thus the calculated spin-flip excitations 
are bounded to the x-ray excited Fe site and do not propagate through the lattice.
Therefore the \textbf{\textit{q}}-dependence is absent in the calculated spectra.
In addition, the spectral weights for the calculated $\Delta S_z=3,~4,~5$ are underestimated with respect to the experimental results by a factor of $\sim$4 (if we take the $\Delta S_z=1, 2$ excitations as a reference for the normalization), see the different scaling factors used in Fig. 3(a) of main text. Possible reasons for this discrepancy are presented in the following. In the LDA+DMFT AIM, the excited Fe atom is coupled to a bath with large gap. A (3$d$) spin-flip in the intermediate state (i.e. during the time evolution between x-ray absorption and emission) can happen due to the exchange with the core-hole or with the bath. However, in the AIM description a spin-flip in the bath amounts to an excitation across the large gap (because the bath is non-interacting). In reality, however, the environment of Fe contains other interacting Fe atoms, i.e., a spin-flip of an excited atom can happen also via an exchange interaction with other Fe atoms involving a much lower energy scale. For this reason, the non-interacting environment considered by the AIM description may lead to underestimating the intensity for the spin-flip excitations. 

\begin{figure*}
        \centering
        {\includegraphics{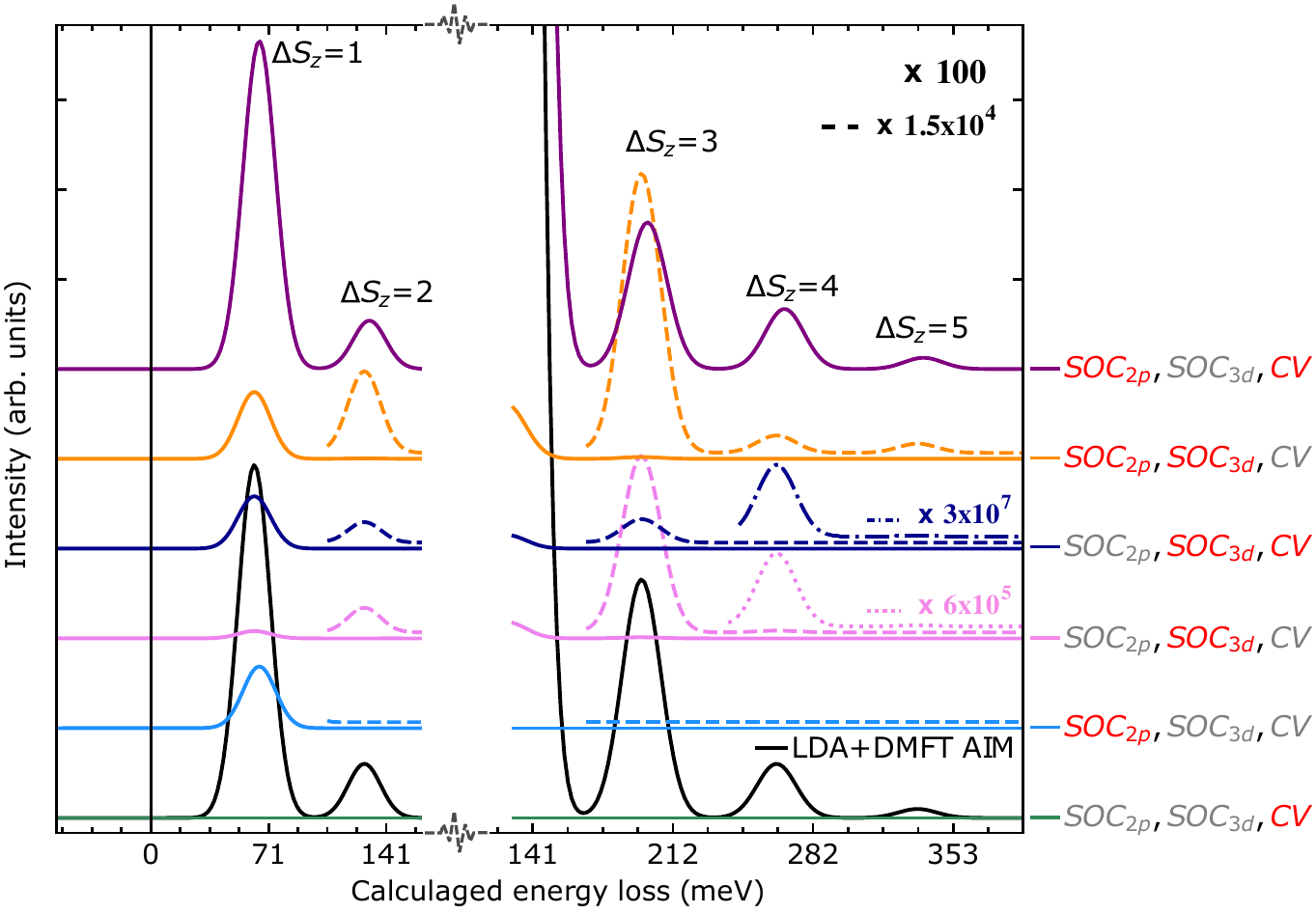}} 
        \caption{The Fe $L_3$-RIXS low-energy features calculated by the LDA+DMFT AIM (black line) and the LDA+DMFT AIM with modified local Hamiltonian (colorful lines). Three parameters: the core-valence (CV$_{2p \leftrightarrow 3d}$) multiplet interaction, the spin-orbital couplings on Fe 2$p$ (SOC$_{2p}$) and 3$d$ (SOC$_{3d}$) shells, are selectively switched on or off in the local Hamiltonian of the AIM to monitor the evolution of multi-spin-flip excitations. The dashed lines are magnified with respect to the solid lines. The red color in the labels indicates the active interaction while the grey color identify the inactivate term.}
        \label{Fig_sm6}
    \end{figure*}

Finally, to understand in details the multi-spin-flip RIXS intensity for the studied $S=\frac{5}{2}$ system, we independently examine the effect caused by each interaction included in our model. The LDA+DMFT AIM results presented in Fig.~3(a) of the main text in fact implement several interactions simultaneously:~1)~the RIXS dipole-dipole interaction, 2)~the Coulomb interaction (including multipole terms) between the Fe 2$p_{3/2}$ core-hole and valence Fe 3$d$ electrons, 3) the exchange field generated by the ordered Fe spins, surrounding the x-ray excited site, which is encoded in the DMFT hybridization intensities, 4) SOC in both Fe 2$p$ and Fe 3$d$ states.
To see how important these interactions are, 
we evaluate the RIXS spectra by switching on/off the multipole part of the CV$_{2p \leftrightarrow 3d}$ interaction, the SOC on the Fe 2$p$ (SOC$_{2p}$) and 3$d$ (SOC$_{3d}$) shell in the AIM Hamiltonian $\hat{H}_{\rm AIM}$. The SOC$_{2p}$ and SOC$_{3d}$ break the  conservation of the spin angular momentum on the Fe 3$d$ and 2$p$ shell, respectively. 
Thus, the multi-spin-flip excitations are in principle possible when the SOC$_{3d}$ is present, see the pink line in Fig.~\ref{Fig_sm6}.
However, since the SOC$_{3d}$ constant is tiny, the spectral intensities of these excitations are negligibly weak. 
The SOC$_{2p}$ constant is about 200 times stronger than SOC$_{3d}$ one, see Table~\ref{t_soc}, and it is responsible for the observed intensities of the multi-spin-flip excitations by opening up multiple paths in the intermediate state (e.g. see yellow line in Fig.~\ref{Fig_sm6}).
Furthermore, the Fe 2$p$ core and 3$d$ valence electrons are coupled with each other by the CV$_{2p \leftrightarrow 3d}$ term, which is relatively strong at the $L$ edge. Thus, the presence of the SOC$_{2p}$ together with the CV$_{2p \leftrightarrow 3d}$ interaction allows the multi-spin-flip excitations on the Fe 3$d$ shell yielding a considerable RIXS intensity for $\Delta S_z=1 - 5$ (see purple line), almost similar to the case where all interactions are switched on (see black line). In fact, when the CV$_{2p \leftrightarrow 3d}$ (and SOC$_{3d}$) is switched off, only $\Delta S_z=1$ is possible, see blue line in Fig.~\ref{Fig_sm6}. This observation is consistent with recent study on double-spin flip of NiO at Ni $L_3$-edge RIXS~\cite{Nag20}.

\jl{In our discussion, we left the valence Coulomb interaction unchanged. The reason is twofold. As the valence Coulomb interaction is the dominant term in $\alpha$-Fe$_2$O$_3$, it determines the high-spin nature of the local ground state. By varying the valence Coulomb interaction in the RIXS initial state, it would alter the ground state of the system, e.g. not S=5/2 anymore, resulting in an “unrealistic” ground state. Additionally, in the RIXS intermediate state, the valence Coulomb interaction does not break the spin-rotational symmetry, therefore conserves the Fe 3d spin. Thus, this term is irrelevant for the presence or absence of the multi-spin-flip excitations in the RIXS process.}

    \begin{figure}
        \centering
        {\includegraphics[width=\linewidth]{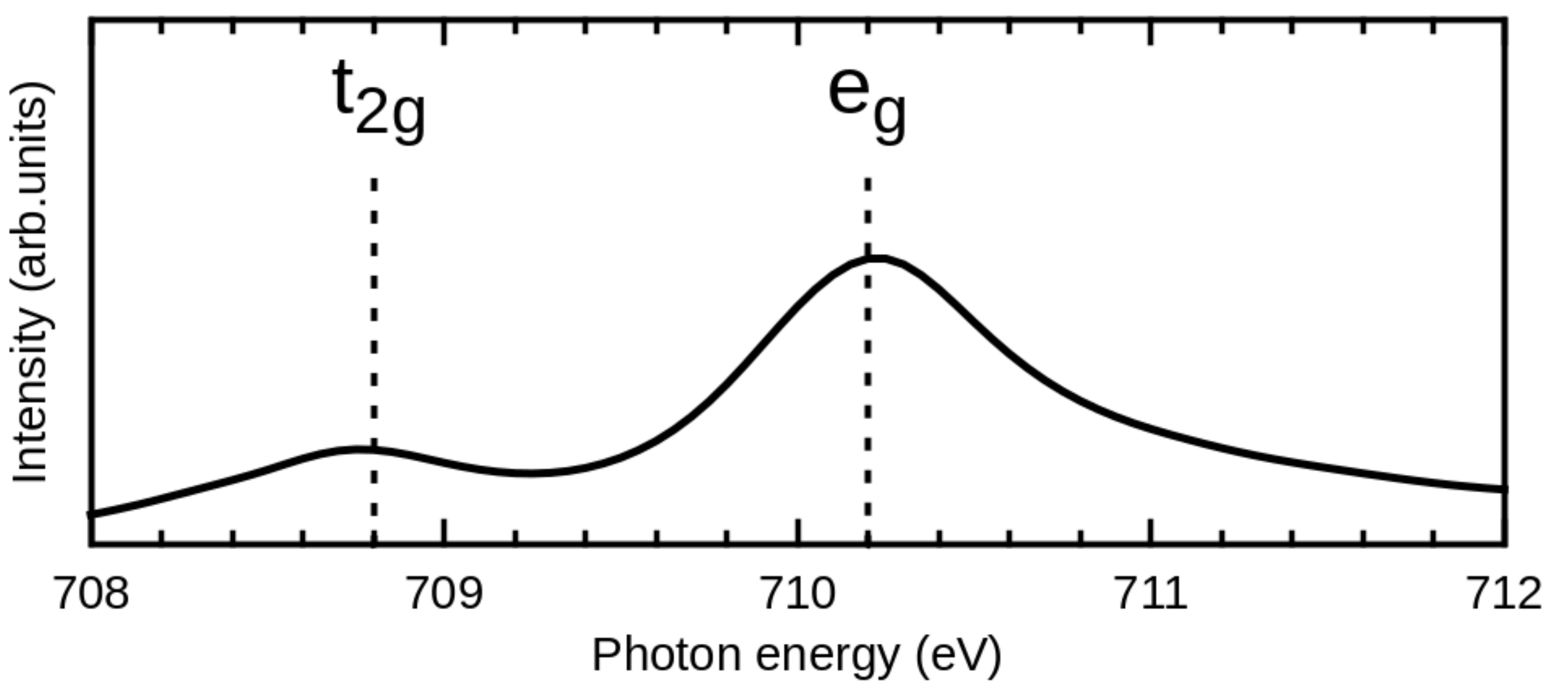}} %
        
        \caption{Calculated Fe $L_3$ XAS spectrum of $\alpha$-Fe$_2$O$_3$ using the LDA+DMFT AIM. The dashed lines indicate the Fe $t_{2g}$ and the $e_g$ resonant energies extracted from the experimental XAS of Fig. 1(b), underlying the good agreement between the calculated and measured XAS spectra. All RIXS calculations were done by setting the incident energy to the Fe $e_g$ resonance, likewise the experiment.}
        \label{Fig_sm7}
    \end{figure} 

        \section{Atomic model simulation for RIXS amplitude}
        In Fig.~4 of the main text, we present the RIXS amplitude 
        ${I_{\rm m}=| \langle f| T_e |m\rangle \langle m | T_i |g\rangle |^2} \notag$
        for specific intermediate states $|m\rangle$
        to understand how the RIXS process can lead locally to multi-spin-flip excitations in this $S=\frac{5}{2}$ system. 
        Computing the amplitude $I_m$ requires the knowledge of the eigenvectors $|m\rangle$ explicitly, thus the LDA+DMFT AIM cannot be used for this purpose.  
        We then examined the computationally feasible atomic model \cite{Groot2021} which was done using in-house code \cite{Wang18} and checked independently using EDRIXS \cite{Wang19}. The two impurity models implement the same local Hamiltonian describing the x-ray excited Fe site (see values in Table~\ref{t_soc}), except for the crystal-field term defining the Fe 3$d$ energy levels.
        The crystal-field term in the atomic model should incorporate contributions from anisotropic hybridization of the Fe 3$d$ orbitals with ligands, and thus must be different from the value in the LDA+DMFT AIM that encodes the hybridization with surrounding ions (bath) explicitly. For the atomic model, the crystal-field parameter is set to $10Dq = $~1.8~eV~\cite{Groot90} which reproduces the Fe $L_3$-XAS shape in the LDA+DMFT AIM result well. We neglected a small trigonal distortion since it is not important for the conclusions drawn in the main text.
        To mimic the exchange field generated by the ordered Fe spins, that is encoded in the hybridization intensities in the LDA+DMFT AIM, we apply a local magnetic field $H_{ex}=-\vec{\boldsymbol{\mu}}_s \cdot \vec{\boldsymbol{B}}$ with $ \boldsymbol{B}=(0.033,0,0)$ in the unit of eV to the atomic model so that the spin-flip excitations $\Delta S_z=1, 2, 3, 4, 5$ cost 66~meV, 132~meV, 198~meV, 264~meV, 329~meV, respectively.


%